\documentclass[pra,twocolumn,superscriptaddress,amsmath,amssymb]{revtex4-2}
\usepackage{bm}
\usepackage{graphicx}
\usepackage{color}
\usepackage{dcolumn,amsmath}
\usepackage[normalem]{ulem}
\usepackage[export]{adjustbox}
\usepackage[caption=false,labelformat=simple]{subfig} 

\begin{document}

\title{Long-range states in collisions of ultracold molecules}

\author{James F. E. Croft}
\affiliation{Joint Quantum Centre (JQC)
Durham-Newcastle, Department of Chemistry, Durham University, South Road,
Durham, DH1 3LE, United Kingdom.}
\author{Brian K. Kendrick}
\affiliation{Theoretical Division (T-1, MS B221), Los Alamos National Laboratory, Los Alamos,
New Mexico 87545, USA}
\author{Jeremy M. Hutson}
\email{j.m.hutson@durham.ac.uk}
\affiliation{Joint Quantum Centre (JQC)
Durham-Newcastle, Department of Chemistry, Durham University, South Road,
Durham, DH1 3LE, United Kingdom.}

\begin{abstract}
We use coupled-channel calculations to explore the nature of near-threshold bound states in a simplified model of Rb+KRb. This is a prototype for systems with very strong coupling at short range and chaotic behavior for the short-range states. We find that there are states with strong long-range character that exist close to threshold and probably persist to depths at least 100 GHz below each threshold. These states are only weakly coupled to the short-range states and do not form part of the chaotic manifold. Since they spend little time at short range, they are relatively insensitive to destruction by laser light. They can thus have long lifetimes that are unrelated to the density of states and can cause narrow Feshbach resonances when the states are shifted across thresholds by external fields.
\end{abstract}
\date{\today}
\maketitle

\section{Introduction}
It is now possible to prepare samples of ultracold molecules and study their collisions with one another and with other collision partners such as alkali-metal atoms~\cite{Ospelkaus:react:2010, Takekoshi:RbCs:2014, Park:NaK:2015, Guo:NaRb:2018, Liu:bimol:2022, Zhao:control:2022}. The collisions can often be studied as functions of external parameters such as electric, magnetic and laser or microwave fields. This allows an unprecedented degree of control and provides new insights into molecular collision dynamics at its most fundamental level. A detailed theoretical understanding is needed to complement this extraordinary experimental capability and to realize the many proposed applications of ultracold molecules.

The alkali-metal diatomic molecules (alkali dimers) form an important class of ultracold molecules. They are usually made by magnetoassociation of a pair of ultracold atoms, followed by stimulated Raman adiabatic passage (STIRAP) to transfer the molecules to their rovibrational ground states~\cite{Ni:KRb:2008, Danzl:ground:2010, Langen:2024}. Once there, the molecules can be manipulated with optical, microwave and magnetic fields to transfer them to selected rotational, hyperfine and Zeeman states \cite{Ospelkaus:hyperfine-control:2010, Gregory:RbCs-microwave:2016, Langen:2024, Cornish:2024}.

There is now much evidence that collisions of ultracold molecules are mediated by complex formation, sometimes known as ``sticky collisions"~\cite{Mayle:2012, Mayle:2013, Ye:2018, Gregory:RbCs-collisions:2019, Bause:2023}. For $^{40}$KRb+$^{40}$KRb, the complexes have been detected directly \cite{Hu:2019}. In many circumstances, complex formation leads to loss of molecules from the trap, even if there is no 2-body reaction pathway available. This is believed to occur via laser excitation of the complexes by the trapping light \cite{Christianen:laser:2019, Gregory:RbCs-collisions:2019, Liu:2020}, though other mechanisms may contribute too.

The nature and properties of the collision complexes remain controversial. It is generally accepted that the strong interactions in both atom-diatom and diatom-diatom systems produce dense sets of levels that show signatures of quantum chaos. These levels and their consequences for collisions were first investigated by Mayle \emph{et al.}~\cite{Mayle:2012, Mayle:2013}, although the densities of states were revised downwards by later work~\cite{Croft:2017, Christianen:density:2019, Frye:triatomic-complexes:2021}.

Chaotic states are generally modeled using random matrix theory. When such states couple to a scattering continuum, they are expected to have widths that are proportional to their mean spacing, as given by Rice-Ramsperger-Kassel-Marcus (RRKM) theory. The densities of states can be used to estimate mean widths and lifetimes, although there are open questions around the distributions of widths~\cite{Croft:lifetimes:2023} and whether to include nuclear spins \cite{Mayle:2013, Jachymski:2022} and the effects of external fields \cite{Man:2022} in the density of states. For some collision systems such as RbCs+RbCs~\cite{Gregory:RbCs-complex-lifetime:2020} and $^{40}$KRb+$^{40}$KRb~\cite{Liu:2020}, measured lifetimes of complexes are in qualitative agreement with the simple estimates. For other systems, however~\cite{Bause:2021, Gersema:2021}, the experiments imply lifetimes that are orders of magnitude larger than the predictions.

A further issue is that most early experiments with ultracold molecules observed fast loss~\cite{Takekoshi:RbCs:2014, Park:NaK:2015, Guo:NaRb-collisions:2018, Ye:2018}, often at the ``universal" rate~\cite{Idziaszek:PRL:2010, Idziaszek:PRA:2010}, implying that almost every colliding pair that reaches short range is lost from the trap. This can be explained if the widths of the states are large compared to their separation, so that almost every collision is resonant. However, this is incompatible with the very long lifetimes (and consequently small widths) observed in refs.~\cite{Bause:2021, Gersema:2021}.

Similar disparities exist between different atom-diatom systems. Here the densities of states are substantially lower \cite{Christianen:density:2019, Frye:triatomic-complexes:2021} and hyperfine interactions are much larger because of the presence of an unpaired electron spin \cite{Frye:triatomic-complexes:2021}. In cases where chemical reactions are energetically allowed, universal loss is usually observed. However, when reactions are energetically forbidden, there is a wider range of behavior: Cs+RbCs shows near-universal loss~\cite{Gregory:atom-molecule:2021}, while for $^{39}$K+Na$^{39}$K~\cite{Voges:2022} the loss is much slower and depends strongly on the atomic hyperfine state. For Rb+$^{40}$KRb, the rate is close to universal yet the directly measured lifetimes of complexes \cite{Nichols:long-lived:2021} are a factor of $10^5$ larger than estimates based on chaos.

The unpaired electron spins of the alkali-metal atoms open up the possibility of magnetically tunable Feshbach resonances in atom-diatom collisions. Different atomic hyperfine sublevels have quite different Zeeman effects, so near-threshold bound states corresponding to one sublevel may be tuned across a threshold corresponding to another. The resulting Feshbach resonances may be used either to control scattering lengths and cross sections (potentially allowing sympathetic cooling) or to form triatomic molecules. The hyperfine structure also opens up many fascinating opportunities for studies of state-to-state collision properties \cite{Liu:2025}. However, the hyperfine and Zeeman Hamiltonians and couplings are far more complicated for alkali-metal atom-diatom systems than for atom-atom systems. A detailed consideration of these effects is beyond the scope of the present work, but ref.\  \cite{Frye:triatomic-complexes:2021} has discussed the hyperfine couplings that are present in these systems, together with issues of angular momentum couplings and selection rules.

A particularly interesting system is $^{40}$K+Na$^{40}$K. In this case tunable Feshbach resonances have been observed, superimposed on fast loss~\cite{Yang:K_NaK:2019, Wang:K_NaK:2021}. Scattering resonances occur due to the interference of incoming and outgoing waves; in the presence of universal loss, there is no outgoing wave, so resonances cannot occur~\cite{Frye:2015}. The patterns of magnetic Feshbach resonances have been interpreted as due to near-threshold states that spend most of their time at long range and are only weakly coupled to the chaotic bath of short-range states, so that their behavior is regular rather than chaotic~\cite{Wang:K_NaK:2021, Frye:long-range:2023}. The resonances have subsequently been used to control elastic scattering rates \cite{Su:elastic:2022}, to form triatomic molecules by magnetoassociation \cite{Yang:magnetoassociation:2022} and to perform photoassociation spectroscopy \cite{Cao:2024}.

A recent development is the observation of electric-field dependence of Feshbach resonances. For $^{39}$K+Na$^{39}$K, Meyer zum Alten Borgloh \emph{et al.} \cite{Meyer:2026} observed that the magnetic fields at which resonances occur can be tuned with electric fields. They interpreted their results in terms of triatomic bound states with free internal rotation of the diatomic component, with some evidence of hindered rotation in one case. These results provide further support for the idea that the resonances are due to long-range states that experience only weak anisotropy from the interaction potential.

The purpose of the present paper is to investigate the possible existence of near-threshold states with predominantly long-range character, using coupled-channel bound-state calculations for a simplified model of Rb+KRb. We show that there are indeed states that spend most of their time at long range and are only weakly coupled to chaotic short-range states. For such long-range states, there is no reason to expect that the widths are related to the density of states. They may be much narrower, as is found for long-range states of Van der Waals complexes \cite{Ashton:1983}, with corresponding longer lifetimes. This may explain some of the puzzling discrepancies between experimental measurements and theoretical estimates of lifetimes, and help explain how long-lived complexes and narrow Feshbach resonances can exist in systems with fast collisional loss.

\section{Theory}

The short-range interaction potentials for alkali-metal atom-molecule and molecule-molecule systems are typically deep and anisotropic. The energy of the dissociation limit corresponds to excitations of tens of diatomic vibrational levels or hundreds of rotational levels. As such it is common to treat the short-range states statistically by defining a density of states $\rho_\textrm{sr}$ at threshold; this can be estimated straightforwardly for a given potential~\cite{Mayle:2012, Mayle:2013, Christianen:density:2019, Frye:long-range:2023}. The short-range dynamics is typically assumed to be chaotic and the short-range states to follow the predictions of random matrix theory. This is supported by evidence from classical trajectory calculations~\cite{Croft:complexes:2014, Klos:2021} and quantum dynamics calculations~\cite{Croft:K+KRb:2017, Croft:2017}.

The usual estimates of densities of states do not account for states that exist near dissociation. These states spend most of their time at long range \cite{LeRoy:1970}, where the potential is very shallow and the couplings are relatively weak. At these distances, only a single vibration-rotation channel contributes significantly. It is an open question whether these states are coupled to the short-range ones strongly enough to form part of the chaotic bath, or exist separately from it.

The interaction potential for an atom-diatom system has long-range form $-C_6/R^6$, where $R$ is the intermolecular distance. Solutions to the Schr\"odinger equation for a single-channel system with a potential of this form are known analytically~\cite{Gao:C6:1998}. This allows the location of near-threshold bound states within a set of ``bins" whose depths are multiples of a system-dependent energy $\bar{E}=\hbar^2/(2\mu\bar{a}^2)$, where $\mu$ is the reduced mass and $\bar{a}=(2\pi/\Gamma(1/4)^2)(2\mu C_6/\hbar^2)^{1/4}$ is the mean scattering length~\cite{Gribakin:1993}. There is then one s-wave state (with relative angular momentum $L=0$) in a ``top bin" between 0 and $39.2\bar{E}$ below threshold, a second in ``bin 2" between $39.2\bar{E}$ and $272.5\bar{E}$, a third between $272.5\bar{E}$ and $871\bar{E}$, and so on. The positions of the states within each bin are determined by the scattering length: for large positive scattering lengths, the states all lie near the top of the bins, while for large negative scattering lengths they all lie near the bottom of bins.

In a multichannel system, a long-range states exists in each bin below every threshold, and states arising from different thresholds may be coupled to one another. For Rb + KRb collisions, $\mu= 51.578\ m_\textrm{u}$ and $C_6= 9010\ E_\textrm{h} a_0^6$. This gives $\bar{a} = 97\ a_0$, which translates to bin boundaries of 134, 927, and 2961~MHz below threshold for the top three states.

\subsection{Potential Energy Surface}
In this work, we use a recently developed potential energy surface (PES) for the RbKRb system that is full-dimensional and is based on a first-principles ({\it ab initio}) methodology. A detailed description of this PES is given in Ref.\ \cite{Kendrick:2026}, so only a brief summary of the most notable features is given here. The surface includes an accurate treatment of the long-range interactions as well as both the ground and first excited electronic states. The global minimum for the ground adiabatic electronic state occurs at a $C_{2v}$ geometry, near $\rho=12.6\,a_0$, and lies 58,126~GHz below the asymptotic energy.

The PESs for the ground and first excited adiabatic electronic states exhibit a conical intersection (degeneracy) along a one-dimensional seam in the three-dimensional nuclear coordinate space that corresponds to $C_{2v}$ geometries (i.e., isosceles triangle configurations for which the two KRb internuclear distances are equal). The global minimum energy of the conical intersection occurs deep within the KRb$_2$ potential well near $\rho=11.5\,a_0$ and lies 41,780~GHz below the asymptotic energy of Rb + KRb($v=0, j=0$).

The conical intersection is energetically accessible within the interaction region and plays an important role in the quantum dynamics at short range \cite{Kendrick:2026}. However, the excited electronic state correlates at long range with a threshold that involves an electronically excited atom and lies 384 THz above that for a ground-state atom. The excited state thus does not contribute substantially to the density of states around the lowest threshold. Moreover, vibrational states supported by it do not extend far to long range, where they might interact with the long-range states of interest here. In the present work we ignore the excited electronic state and perform full-dimensional quantum dynamics calculation on the ground adiabatic electronic state.

Near-threshold states are extremely sensitive to variations in the potential. No PES from \emph{ab initio} calculations could be accurate enough to give specific predictions for them. In this work, therefore, we vary the PES by small amounts around the one of ref.\ \cite{Kendrick:2026} using a scaling factor $\lambda$, as described below. This allows us to obtain insights that are independent of any specific PES.

\subsection{Hamiltonians and basis sets}
We perform calculations both in full dimensionality, using a hyperspherical approach, and in reduced dimensionality with an atom + rigid-rotor approach. In both calculations we fix the total angular momentum $J$ to zero and neglect the role of nuclear and electronic spins and of any non-adiabatic effects due to the excited electronic surface. The restriction to $J=0$ is appropriate for the states sampled in ultracold experiments, which usually involve non-rotating molecules and are at collision energies in the s-wave regime.

The Hamiltonian for a nonrotating three-particle system in hyperspherical coordinates is
\begin{equation}
  H = -\frac{\hbar^2}{2\mu_3\rho^5}\frac{\partial}{\partial\rho}\rho^5\frac{\partial}{\partial\rho} + \frac{\hat{\Lambda}^2}{2\mu\rho^2} + \hat{V},
  \label{eq:hamhyp}
\end{equation}
where $\rho$ is the hyperradius, $\mu_3=[m_\textrm{A}m_\textrm{B}m_\textrm{C}/(m_\textrm{A} + m_\textrm{B} + m_\textrm{C})]^{1/2}$ is the three-body reduced mass, $\hat{\Lambda}$ is the grand angular momentum operator, and $\hat{V}$ is the potential-energy operator. We use adiabatically adjusting principle-axis hyperspherical (APH) coordinates, which treat all three arrangement channels fully equivalently~\cite{PACK:1987, Kendrick:1999}.
We expand the total wavefunction $\Phi$ as
\begin{equation}
\Psi = \rho^{-5/2} \sum_i \Phi_i(\xi) \chi_i(\rho),
\end{equation}
where $\xi$ represents all coordinates except $\rho$ and the internal basis functions $\Phi_i(\xi)$ are 5D APH surface functions as defined in ref.~\cite{Kendrick:1999}. Substituting this expansion into the total Schr\"odinger equation yields the coupled-channel equations, which are a set of 2nd-order differential equations in $\rho$.

The basis set of APH surface functions for hyperspherical calculations, is obtained by diagonalising the Hamiltonian to obtain quasiadiabatic eigenvectors as described in ref.\ \cite{Kendrick:2026}. In the present work, the diagonalization was carried out on a logarithmic grid with sectors centered at 641 different values of $\rho$ between 7.0 and $170~a_0$. In each sector, the lowest 2,250 surface functions of even exchange symmetry were retained for use in the coupled-channel calculation described below; this is sufficient to include all locally open channels and at least 30 locally closed channels.

The Hamiltonian for an atom + rigid-rotor system in Jacobi coordinates is
\begin{equation}
  H = -\frac{\hbar^2}{2\mu R}\frac{d}{dR^2}R + b\hat{n}^2
      +\frac{\hbar^2}{2\mu}\frac{\hat{L}^2}{R^2} + \hat{V}
      \label{eq:hamjac}
\end{equation}
where $\mu=m_\textrm{AB}m_\textrm{C}/(m_\textrm{AB} + m_\textrm{C})$ is the two-body reduced mass, $b$ is the rotational constant of the diatom AB, $\hat{n}$ is its rotational angular momentum operator, and $\hat{L}$ is the angular momentum operator for relative motion. We expand the total wavefunction $\Phi$ as
\begin{equation}
\Psi = R^{-1} \sum_i \Phi_i(\xi) \chi_i(R),
\end{equation}
where $\xi$ now represents all coordinates except $R$ and the internal basis functions $\Phi_i(\xi)$
are formed by coupling $n$ to $L$ to give a total angular momentum $J$. For $J=0$, all basis functions have $L=n$. Substituting this expansion into the total Schr\"odinger equation yields a set of coupled-channel equations in $R$.

In the present work, the basis set for atom + rigid-rotor calculations includes all KRb rotor functions up to $n_\textrm{max}=220$, which gives good convergence; only 137 channels are ever locally open.

\subsection{Solution of coupled-channel equations}

We solve the coupled-channel equations to find bound-state solutions \cite{Hutson:bound:1984}. The general strategy is to propagate separate multichannel solutions of the Schr\"odinger equation from short and long range, subject to bound-state boundary conditions, to a central point at $R_\textrm{match}$ or $\rho_\textrm{match}$. Bound states are located at energies where the two wavefunctions and their derivatives match at the central point.

For coupled-channel calculations in the atom + rigid-rotor representation, we use the BOUND package \cite{bound+field:2019, mbf-github:2023}. We use the fixed-step log-derivative propagator of Ref.\ \cite{Manolopoulos:1986} from $R_\textrm{min}=5.0\ a_0$ to $R_\textrm{match}=9.03\ a_0$ and the variable-step Airy propagator of Ref.\ \cite{Alexander:1987} from $R_\textrm{max}=400.0\ a_0$ to $R_\textrm{mid}=20.0\ a_0$, followed by the fixed-step log-derivative propagator from $R_\textrm{mid}$ to $R_\textrm{match}$. We calculate wavefunctions by back-substituting into the log-derivative solutions as described in Ref.\ \cite{Thornley:1994}.

For calculations in the hyperspherical representation, we use a program based on the renormalized Numerov propagator, similar to that used for scattering calculations \cite{Kendrick:2026}. We use 100 steps to propagate across each sector. We initially attempted to locate bound states using the methods of ref.~\cite{Kendrick:HO2:1997}, which are based on Johnson's approach of locating zeroes of the determinant of the renormalized Numerov matching matrix \cite{Johnson:1978}. However, this proved difficult to converge with the very large number of closed channels in the present work. We therefore implemented the method described in ref.\ \cite{Hutson:CPC:1994}, which is based instead on locating zeroes of individual eigenvalues of the matching matrix.

For coupled-channel calculations in both representations, a key concept is the multichannel node count \cite{Johnson:1978}. This is defined at each energy $E$ as the number of bound states that lie below $E$. It can be calculated as a by-product of the propagation. It allows the program to locate bound-state energies approximately by bisection before converging on them, and also to ensure that no bound states are missed. In the present work, we used an extension of the node-count algorithm \cite{LeSueur:2026} that identifies the eigenvalue of the matching matrix corresponding to the desired eigenvalue over the full range of energy that it exists. This minimizes the number of bisections needed before switching to a fast root-finding algorithm; it improves the efficiency and reliability of convergence, particularly in systems with large numbers of coupled channels, as in this work.

\section{Results}
\subsection{Adiabats}

\begin{figure}[tb]
\centering
\includegraphics[width=1\columnwidth]{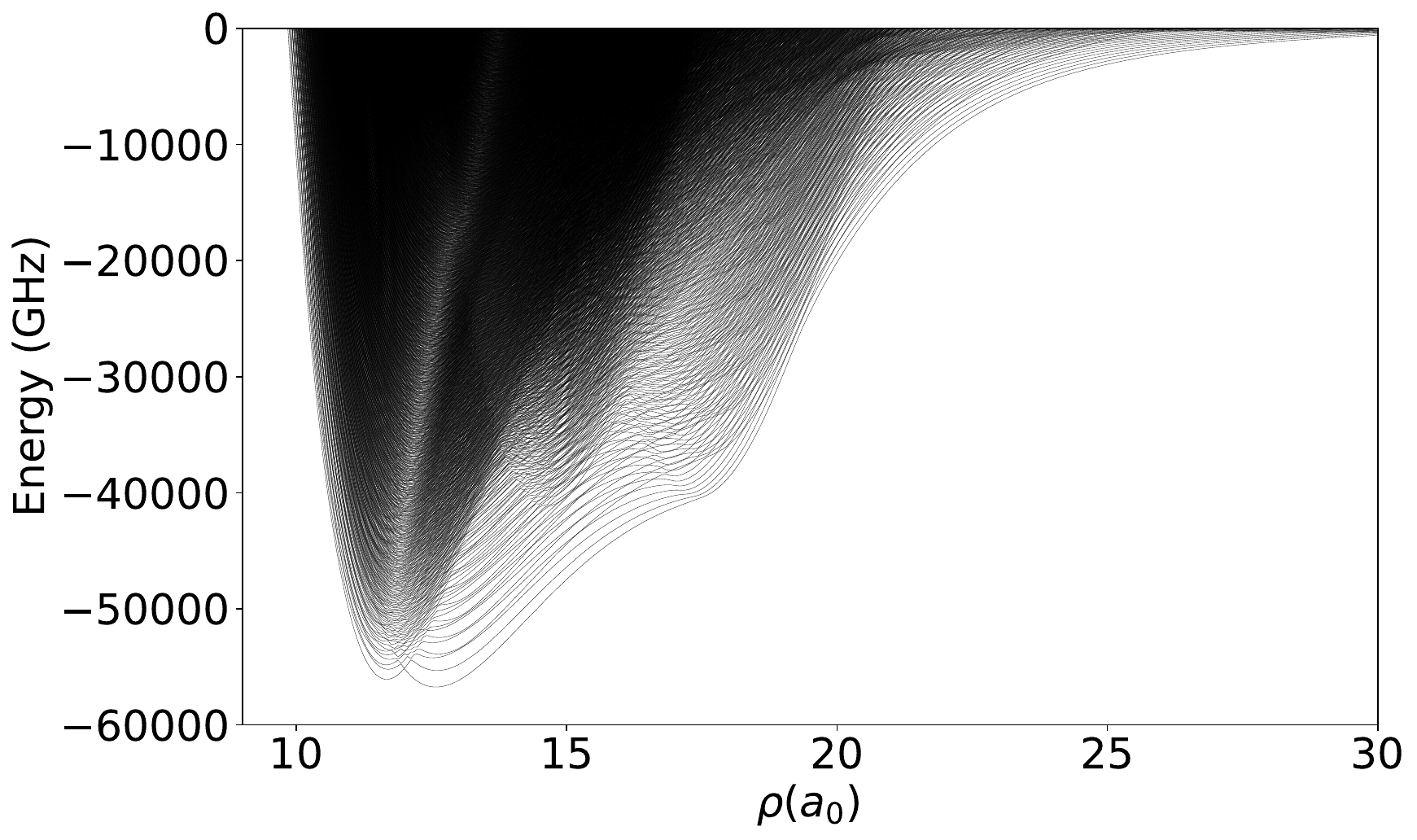}
\includegraphics[width=1\columnwidth]{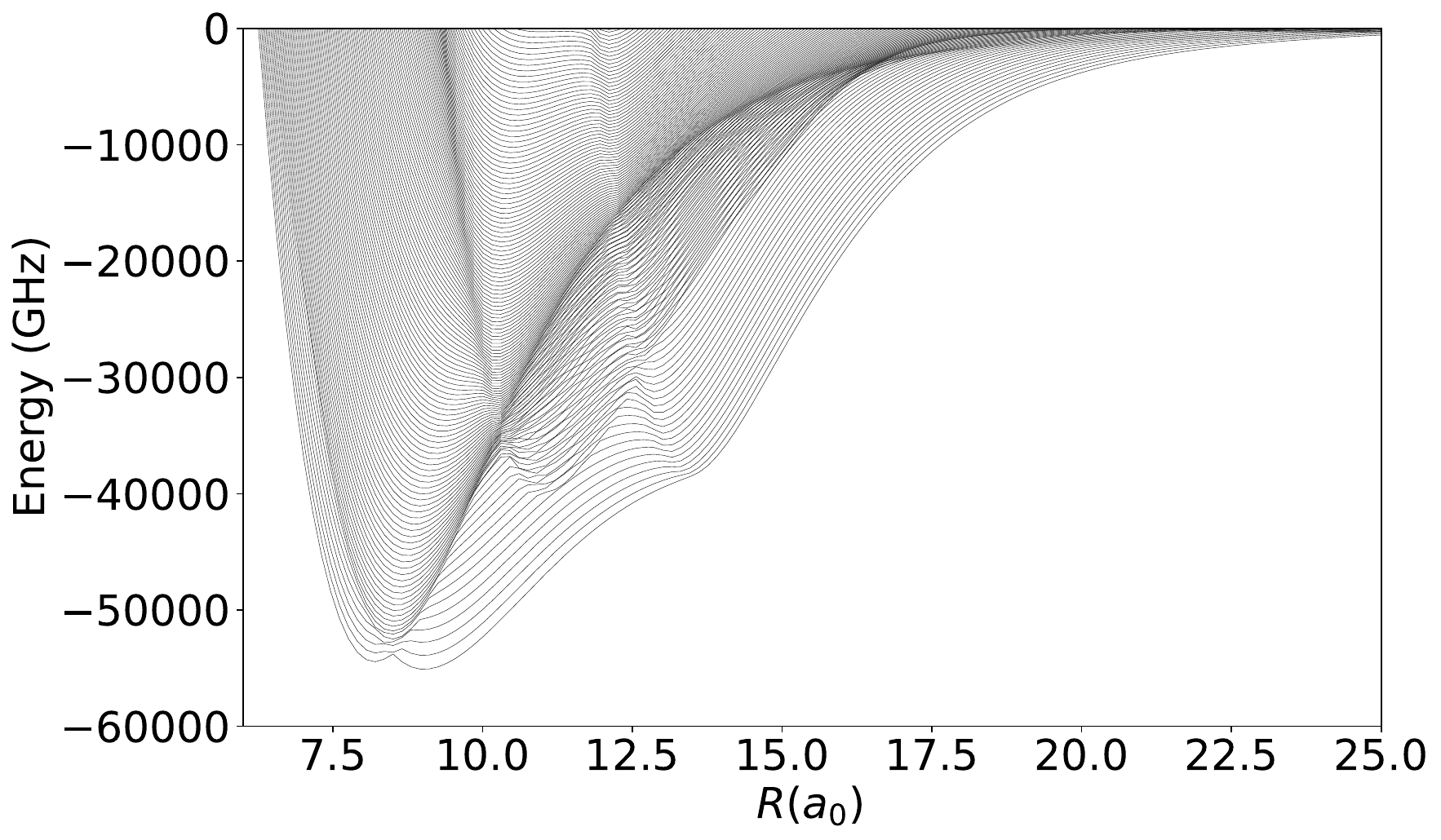}
\caption{The adiabats for Rb+KRb (a) in the hyperspherical representation, as a function of hyperradius $\rho$; (b) in the Jacobi representation, as a function of atom-diatom distance $R$.}
\label{fig:adiabats}
\end{figure}

Many aspects of coupled-channel calculations are best interpreted in terms of adiabats, which are the eigenvalues of the Hamiltonians (\ref{eq:hamhyp}) and (\ref{eq:hamjac}) at fixed values of $\rho$ or $R$, respectively, neglecting the radial kinetic energy. Figure \ref{fig:adiabats}(a) shows the adiabats
obtained by diagonalizing the Hamiltonian of Eq.\ \ref{eq:hamhyp} at each $\rho$, while Fig.\ \ref{fig:adiabats}(b) shows those obtained by diagonalizing the Hamiltonian of Eq.\ \ref{eq:hamjac} at each $R$. The two panels are scaled horizontally to give qualitative correspondence, although the definition of $\rho$ makes this approximate. They show similar features. There are several sets of wells: the outermost is at around $R= 14~a_0$ and $\rho = 18~a_0$ and corresponds to linear Rb-K-Rb geometries; another set with minima at around $R = 9~a_0$ and $\rho = 14~a_0$ corresponds to linear Rb-Rb-K geometries. The set with minima near $R = 11~a_0$ and $\rho = 12.6~a_0$ correspond to the global minimum at $C_{2v}$ geometries, while the slightly shallower set near $R = 8~a_0$ and $\rho = 11.7~a_0$ correspond to the lower-symmetry bent geometries described in ref.\ \cite{Kendrick:2026}. The comparison between Figs.\ \ref{fig:adiabats}(a) and \ref{fig:adiabats}(b) shows that the atom + rigid-rotor calculations capture much of the physics of the full-dimensional problem, though without some structure due to vibrational excitation of KRb that is visible in Fig.\ \ref{fig:adiabats}(a).

The arrangement channel for K + Rb$_2$($v=0, j=0$) lies 5,680~GHz above the asymptotic energy for Rb + KRb($v=0,j=0$) and is therefore closed for ultracold collisions of Rb with KRb.

\subsection{Node counts and densities of states}
\label{sec:dos}

Our coupled-channel bound-state calculations in hyperspherical coordinates were successful for low-lying states, but we encountered difficulties for near-threshold states. In particular, the precise bound-state energies and wavefunctions obtained were different for different choices of the matching distance $\rho_\textrm{match}$. We believe this arises because the internal basis sets in different sectors span slightly different spaces. As a result, the overlap matrices between them are subunitary, which results in loss of flux at sector boundaries. This is particularly severe in regions where high-lying adiabats of one character (centered around a particular potential minimum) have sharp avoided crossings with adiabats of a different character, as is seen in Fig.\ \ref{fig:adiabats}(a).

\begin{figure}[tb]
\centering
\includegraphics[width=1\columnwidth]{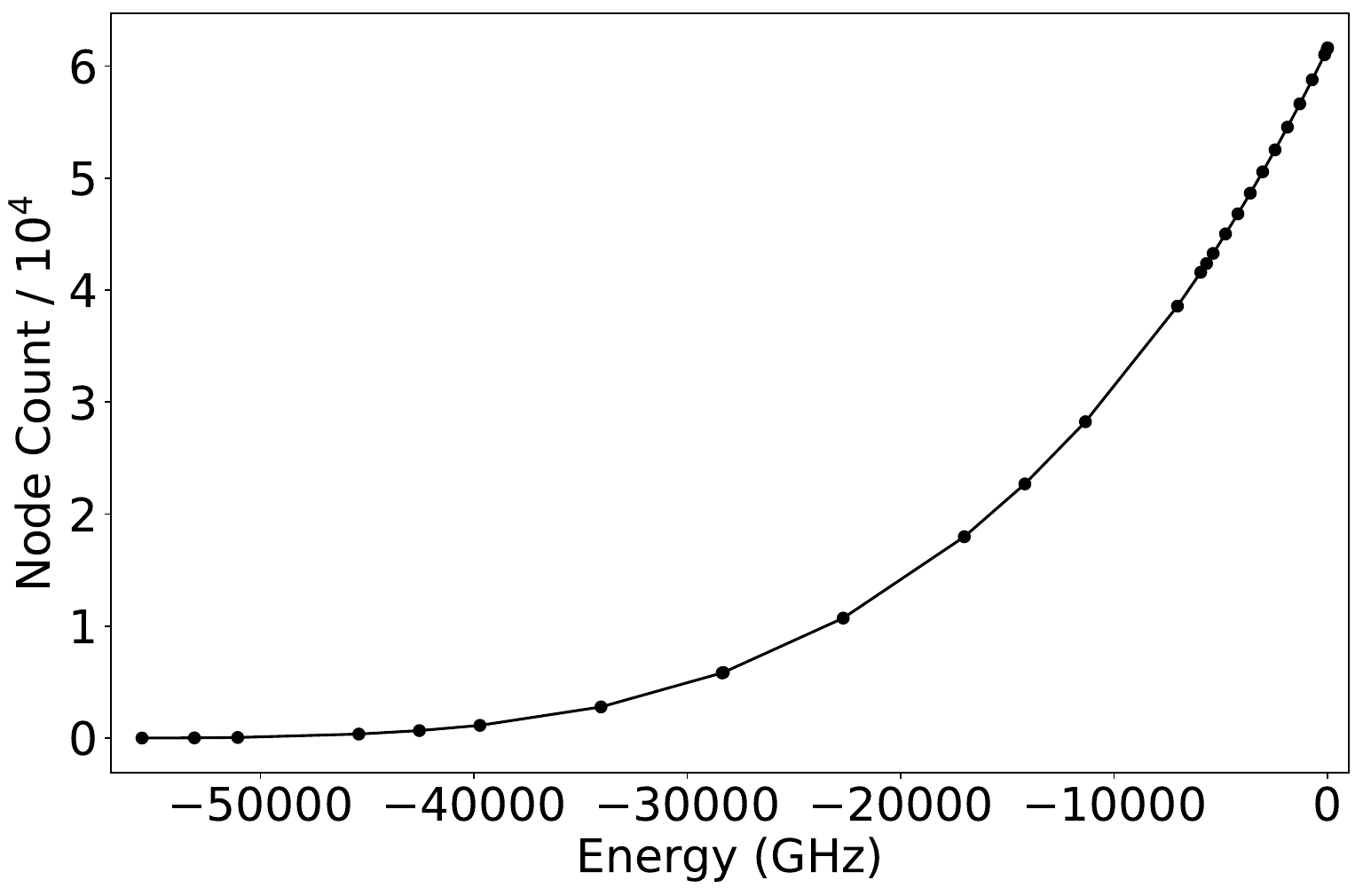}
\includegraphics[width=1\columnwidth,right]{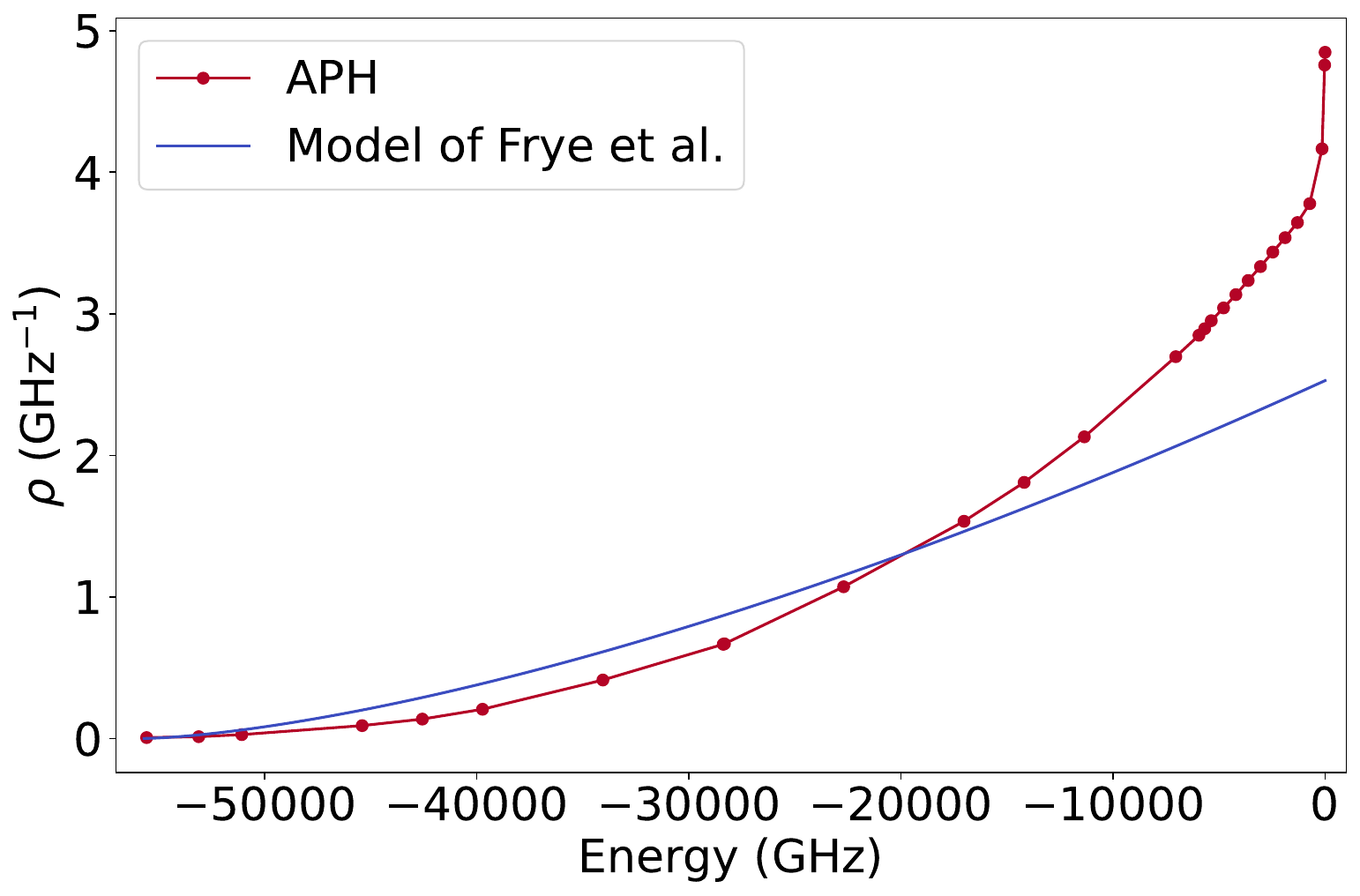}
  \caption{The node count and density of states $\rho$ from hyperspherical coupled-channel calculations (symbols and black/red lines), as a function of energy relative to threshold. The blue line shows the density of states from the model of Ref.\ \cite{Frye:triatomic-complexes:2021}, divided by 2 to count only states of even exchange symmetry.}
\label{fig:dos}
\end{figure}

Despite these difficulties, we obtain a multichannel node count as a function of energy that is reliable to within a few nodes. As described above, this corresponds to the number of bound states that lie below the energy of the propagation. The node count is shown as a function of energy in Fig.\ \ref{fig:dos}(a). Differentiating it yields an accurate density of states that can be compared with models in common use. Figure~\ref{fig:dos}(b) compares the accurate density of states from coupled-channel calculations with the model of ref.\ \cite{Frye:triatomic-complexes:2021}, which is analogous to that of ref.\ \cite{Christianen:density:2019} for molecule-molecule complexes. The model is based on a greatly simplified representation of the potential, harmonic in $R$ and $r$ and independent of $\theta$. The numerical results from the model are divided by 2 in Fig.\ \ref{fig:dos} because the hyperspherical calculations count only states of even Rb exchange symmetry. It may be seen that the real density of states is lower than the model near the minimum, where the model neglects the angular confinement present in the real system. However, the real density of states is substantially higher than the model near threshold. This arises because the harmonic potential used in the model effectively restricts the states to values of $R$ less than about 11~$a_0$. The coupled-channel calculations, by contrast, take full account of states that sample larger values of $R$ or $\rho$.

The derivative of the coupled-channel node count at threshold gives a mean level spacing, $d$, of 206~MHz. This compares to 276~MHz for all states with the model of ref.~\cite{Frye:triatomic-complexes:2021}, corresponding to 552 MHz for states of even exchange symmetry. Figure~\ref{fig:dos}(b) shows that this rather large discrepancy is specific to the region near threshold; it arises because the long-range tail of the potential provides a substantial extra volume of phase space at these energies. The states that occupy this additional phase space may be expected to have long-range character.

\subsection{Extent of quantum chaos}
\label{sec:extent}

As described above, we encountered difficulties in obtaining consistent bound-state eigenvalues and wavefunctions near threshold from hyperspherical calculations. In the remainder of this paper, we therefore use atom-diatom calculations based on the Hamiltonian of Eq.\ \ref{eq:hamjac}. These calculations do not encounter problems due to non-unitary transformations at sector boundaries, because the full basis set is included at every step.

We first verify that the atom + rigid-rotor model reproduces the chaotic dynamics of the full problem, by analysing the bound-state spectrum for signatures of quantum chaos. In the framework of random matrix theory, quantum chaos manifests itself in the statistical distribution of energy levels \cite{Wigner:statisticalnuclear:1951, Dyson:1962}. For a generic non-chaotic system, the scaled nearest-neighbor spacings, $s$, follow a Poisson distribution $P_\mathrm{P}(s) = \exp{(-s)}$~\cite{Berry:1977}. For a chaotic system, on the other hand, they follow a Wigner-Dyson distribution
\begin{equation}
  P_\mathrm{WD}(s) = \frac{\pi}{2}s\exp{\left(-\frac{\pi s^2}{4}\right)}.
\end{equation}
The characteristic difference between these two distributions is in the probability of small nearest-neighbor spacings: the Poisson distribution peaks at a nearest-neighbor spacing of zero, while the Wigner-Dyson distribution exhibits strong level repulsion and drops to zero at zero spacing. In real physical systems, the distribution is often intermediate between the two. To quantify this, it is common to interpolate between them using the Brody distribution \cite{Brody:1973, Brody:rmtreview:1981},
\begin{equation}
\begin{split}
P_\mathrm{B}(s) &= As^\eta\exp{(-\alpha s^{\eta+1})}, \\
A &= (\eta + 1)\alpha, \\
\alpha &= \Gamma\Big(\frac{\eta+2}{\eta+1}\Big)^{\eta+1}.
\end{split}
\label{eq:brody}
\end{equation}
The Brody distribution reduces to $P_\mathrm{P}(s)$ for $\eta=0$ and to $P_\mathrm{WD}(s)$ for $\eta=1$.
The Brody parameter $\eta$ does not have a direct physical interpretation, but Brody parameters less than 1 often arise when a set of chaotic levels (with a Wigner-Dyson distribution) is combined with another set of levels that do not interact with the first \cite{Berry:semiclassical:1984}. Since there is no level repulsion between the two sets, nearest-neighbor spacings close to zero become more probable.

\begin{figure}[tb]
\centering
\includegraphics[width=1\columnwidth]{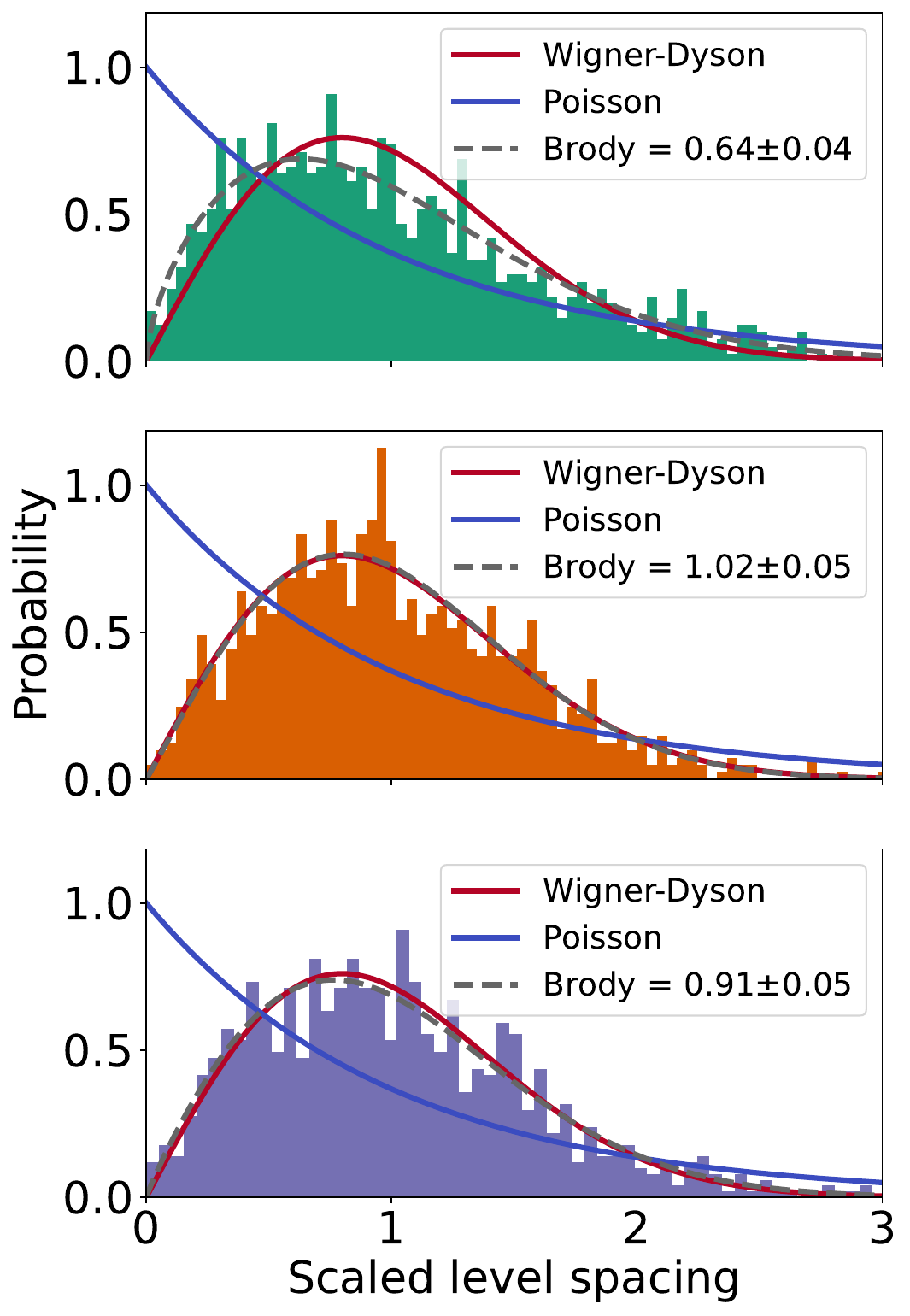}
\caption{The distribution of nearest-neighbor spacings for groups of 1000 bound states near the (a) top, (b) center and (c) bottom of the well, compared with the Wigner-Dyson, Poisson and Brody distributions.}
\label{fig:brody}
\end{figure}

We compute the bound-state spectrum within the atom + rigid-rotor model for $J=0$ and obtain 7288 states in total. Figure~\ref{fig:brody} shows the distribution of nearest-neighbor spacings, shown separately for the top 1,000 bound states (top panel), a set of 1000 states with node counts near 3,600 (middle panel), and the bottom 1,000 states (bottom panel). To remove any variation in the local density of states, we ``unfold'' the energies~\cite{Guhr:1998}. The Brody parameter is then computed by non-linear least-squares fitting of Eq.~\ref{eq:brody} to the energy spacings. For the middle set of states, about two-thirds of the way up the well, it is $1.02 \pm 0.05$, consistent with a quantum-chaotic system. For the states at the bottom of the well, it is a little lower, $0.91 \pm 0.05$, reflecting the regular harmonic nature of bound states localized near the bottom of the well. The 1000 states nearest to threshold, however, yield a substantially lower Brody parameter of $0.64 \pm 0.04$. This is similar to the value of 0.78 obtained for the nearest-neighbor spacings of resonances above threshold in K$_2$ + Rb collisions~\cite{Croft:2017}.

\begin{figure}[tb]
\centering
  \includegraphics[width=1\columnwidth]{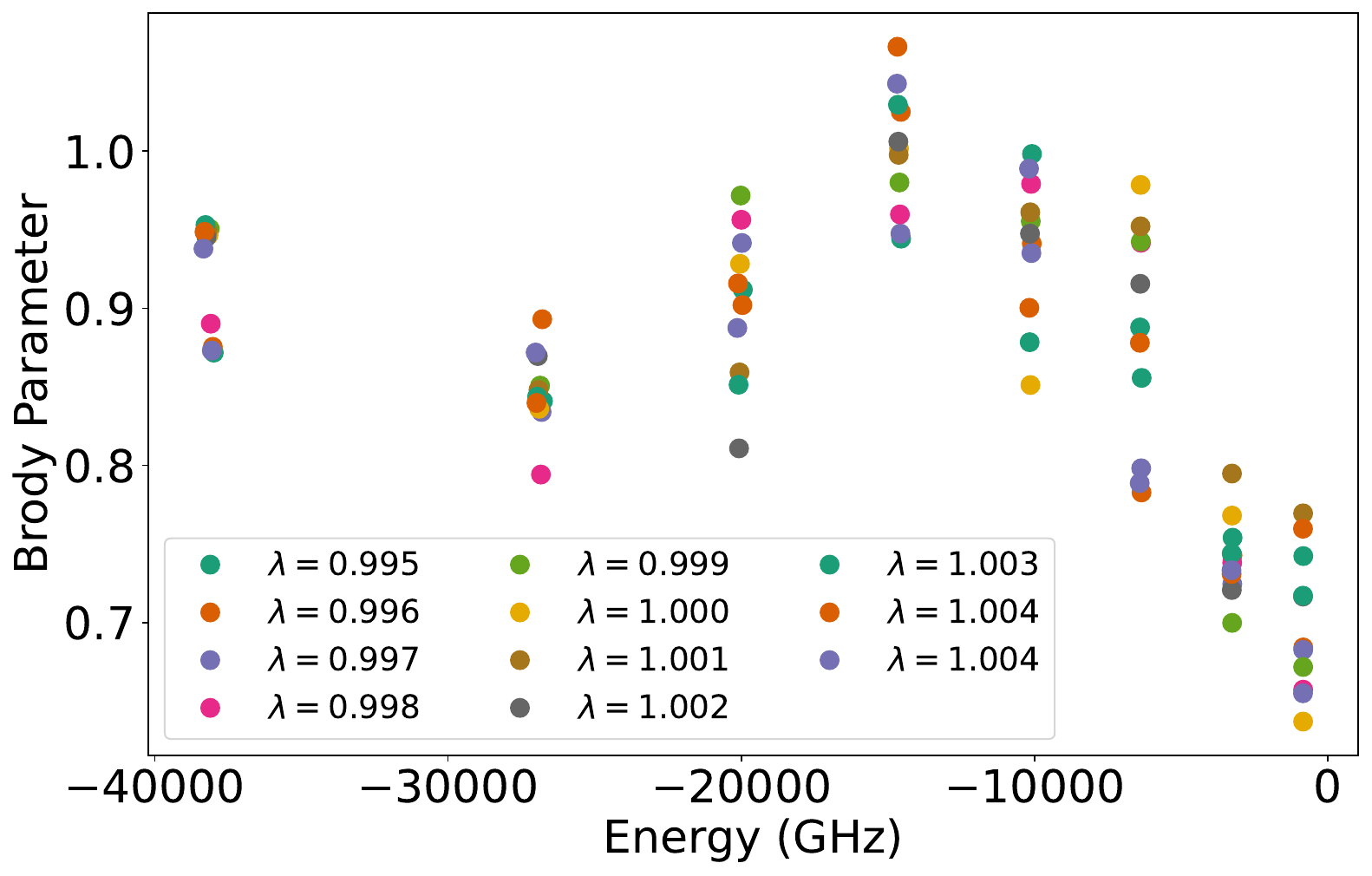}
\caption{The variation of the Brody parameter $\eta$ with energy for potentials with 11 different scaling factors, demonstrating the robustness of the decrease in $\eta$ towards threshold.}
  \label{fig:brody_scaling}
\end{figure}

To verify the robustness of the effect seen in Fig.\ \ref{fig:brody}, we repeat the calculations for different potentials, each multiplied by a scaling factor $\lambda$ between 0.995 and 1.004. This range of scaling factors changes the number of bound states by 83.
For each potential, we partition the bound states into 8 silos of almost equal size and compute the Brody parameter $\eta$ in each silo. Figure~\ref{fig:brody_scaling} plots $\eta$ against the average bound-state energy in the silo concerned. It is clear that the trend seen in Fig.\ \ref{fig:brody} is robust. The Brody parameter increases from the bottom of the well to a maximum near 1 at around $-15$ THz before dropping to around 0.7 near threshold.

It is evident from these results that quantum chaos for Rb+KRb develops with excitation from the potential minimum, and is essentially complete at an energy about 40 THz (1300 cm$^{-1}$) above it. Above that, however, the extent of quantum chaos decreases. This suggests the existence of bound states nearer to threshold that interact only weakly with the short-range states. We investigate this further below.

\subsection{Near-threshold states}

Section \ref{sec:dos} has demonstrated that there are states supported by the long-range tail of the interaction potential. These states exist in addition to the short-range states counted by common models of densities of states, such as those of refs.\ \cite{Christianen:density:2019, Frye:triatomic-complexes:2021}. A key question is the extent to which these long-range (near-threshold) states are coupled to and mix with the short-range chaotic states. To understand this, we examine the behavior of both types of state close to threshold using coupled-channel calculations.

\subsubsection{States in the top bin}

\begin{figure}[tb]
\centering
\includegraphics[width=1\columnwidth]{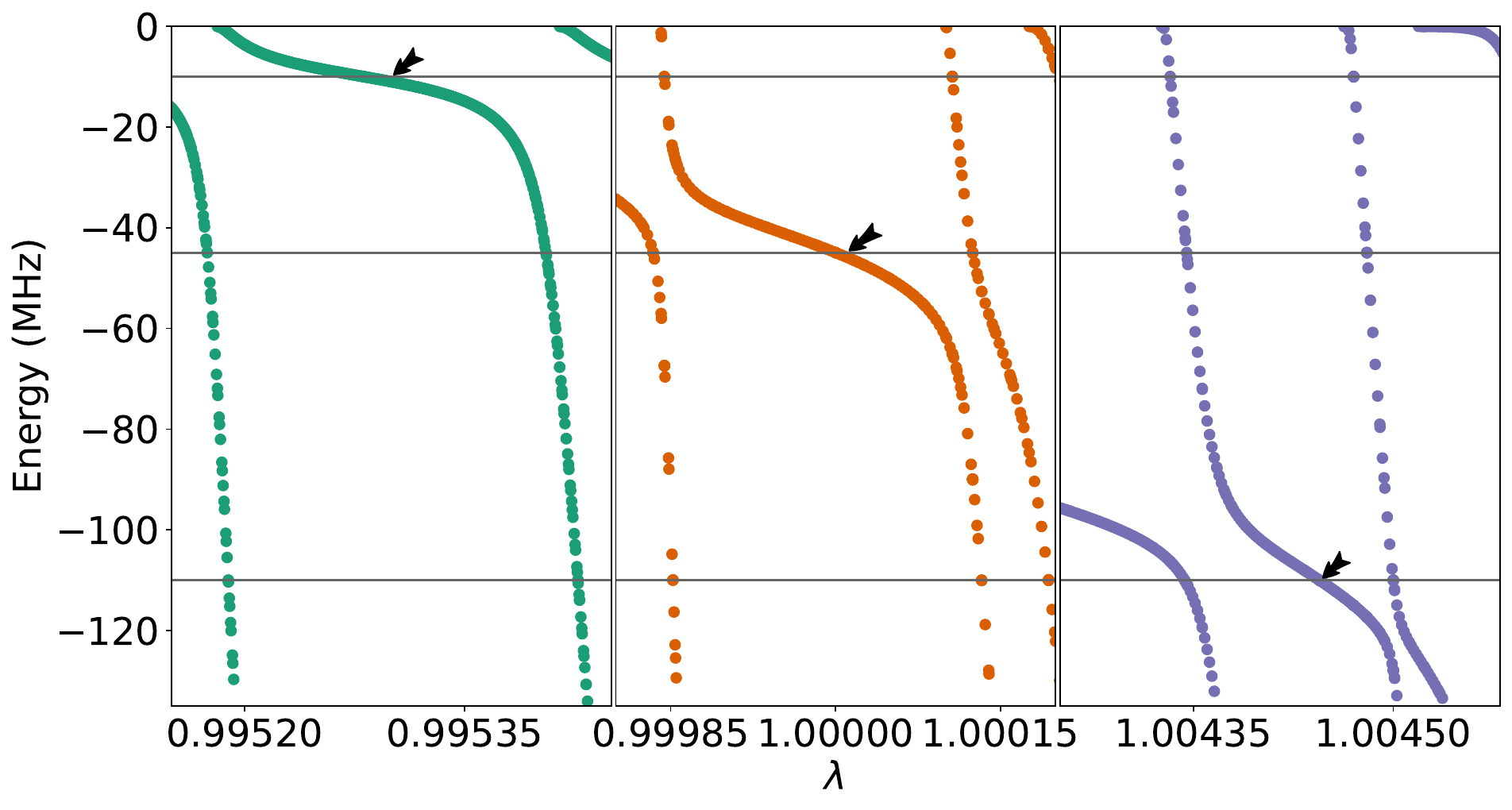}
  \caption{Bound-state energies for states in the top bin, as a function of the scaling parameter $\lambda$. In each panel there is an underlying state whose energy varies slowly with $\lambda$, undergoing avoided crossings with two much steeper states. for values close to $\lambda=1$. The regions of $\lambda$ are chosen to place a long-range state high in the bin (left-hand panel), near the middle of the bin (center panel) and deep in the bin (right-hand panel). The arrows identify the states with wavefunctions shown in Fig.\ \ref{fig:wavefunctions}.}
\label{fig:scaling}
\end{figure}

Figure~\ref{fig:scaling} show the energies of the bound states in the top bin as a function of the scaling factor $\lambda$. Increasing $\lambda$ changes the scattering length and shifts the states downwards. For a single-channel system with $N$ bound states, increasing the potential by a factor $(1+N)^2$ is sufficient to move a near-threshold state all the way from its position in one bin to the corresponding position in the next bin down \cite{Hutson:theory-cold-colls:2009}. States that are supported by the lowest threshold are in narrow bins, so shift downwards relatively slowly as a function of $\lambda$. However, states at the same energy that are supported by higher thresholds are in deeper (and wider) bins with respect to the threshold that supports them, so that they shift downwards faster. In general, the energies of states supported by higher and higher thresholds show progressively larger gradients as a function of $\lambda$.

Across the three panels of Fig.\ \ref{fig:scaling}, a single slowly-varying state that is mostly in the lowest channel, $n=0$, descends through the bin as $\lambda$ increases by about 0.016. The slowly-varying bound state is crossed by many faster-moving states that are supported by higher thresholds with $n>0$, with avoided crossings between them. The three horizontal lines shown in Fig.\ \ref{fig:scaling} are at $-10$, $-44.95$, and $-110$~MHz. The slowly-varying state crosses $-10$~MHz when $\lambda = 0.995$, $-44.95$~MHz when $\lambda = 1$, and $-110$~MHz when $\lambda = 1.004$. For an interaction potential that varies at long range as $-C_6/R^6$, purely long-range states are equally spaced as a function of the cube root of energy~\cite{LeRoy:1970}; accounting for this, the horizontal lines in Fig.\ \ref{fig:scaling} may be viewed 42, 69, and 94\% of the way down the top bin. As the slowly-varying state crosses the bottom of the first bin, a new state appears at the top. This is as expected for a long-range state.

For a single potential curve, the number of bound states is given semiclassically by $N=\Phi/\pi-\frac{1}{2}$, where the phase integral $\Phi$ is
\begin{equation}
\Phi=\int \left( \frac{2\mu [E-V(R)]}{\hbar^2} \right)^\frac{1}{2}\,dR,
\label{eq:phase-int}
\end{equation}
where $E$ is the energy at dissociation and the integral is over the classically allowed region. For scaled potentials, $\Phi$ is proportional to $\lambda^{1/2}$. The observed cycle length of 0.016 in $\lambda$ thus implies that the near-threshold states are supported by an effective potential with $N+\frac{1}{2}=\Phi/\pi\approx 125$. We calculate $\Phi/\pi=147$ for the lowest adiabat by numerical integration of Eq.\ \ref{eq:phase-int}. The difference between these two values indicates that the long-range states do not sample the entire phase space accessible to the lowest adiabat. This is probably due to nonadiabatic effects that prevent the long-range states following the lowest adiabat all the way to its inner turning point.

\begin{figure}[tb]
\centering
\includegraphics[width=0.9\columnwidth]{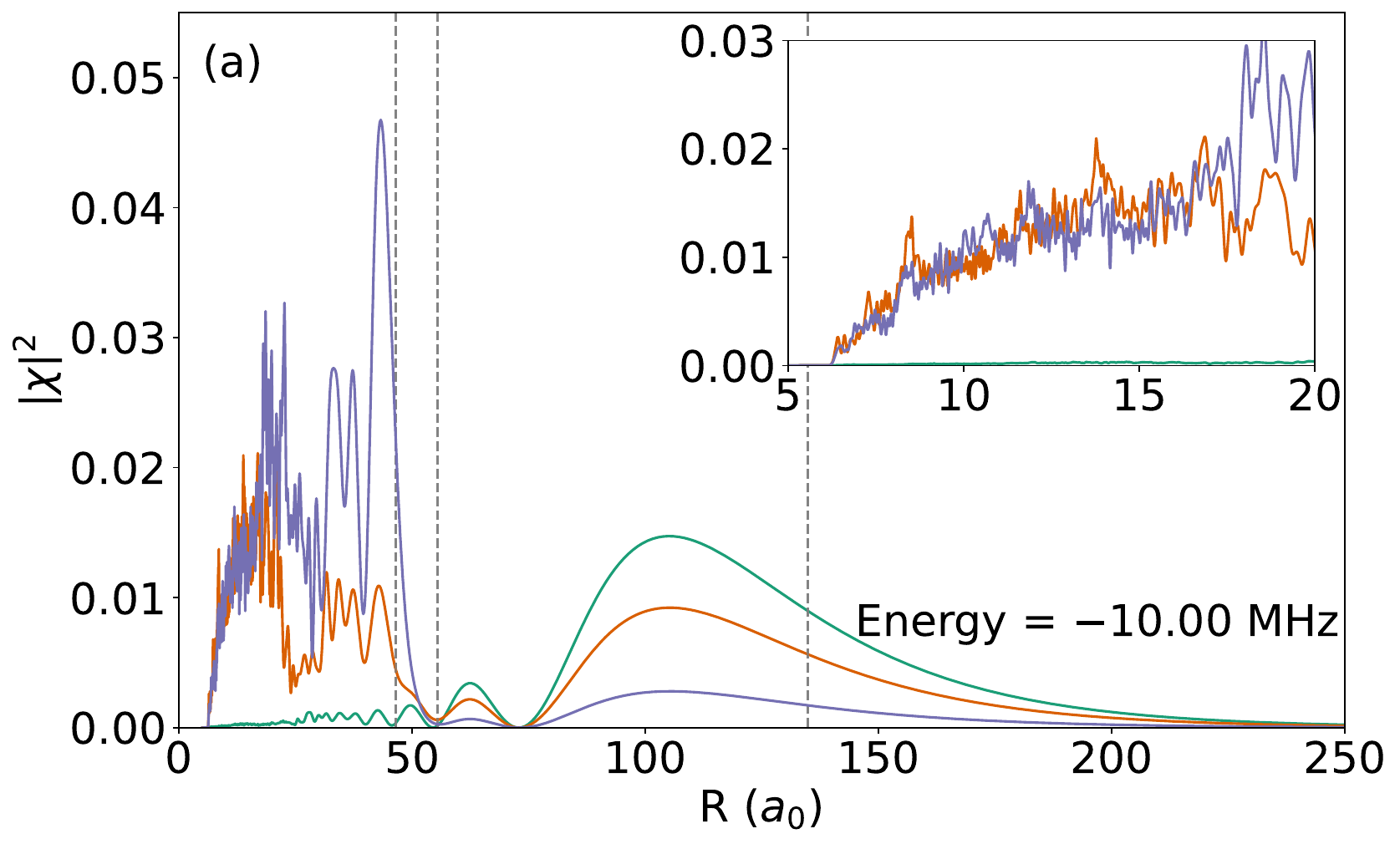}
\includegraphics[width=0.9\columnwidth]{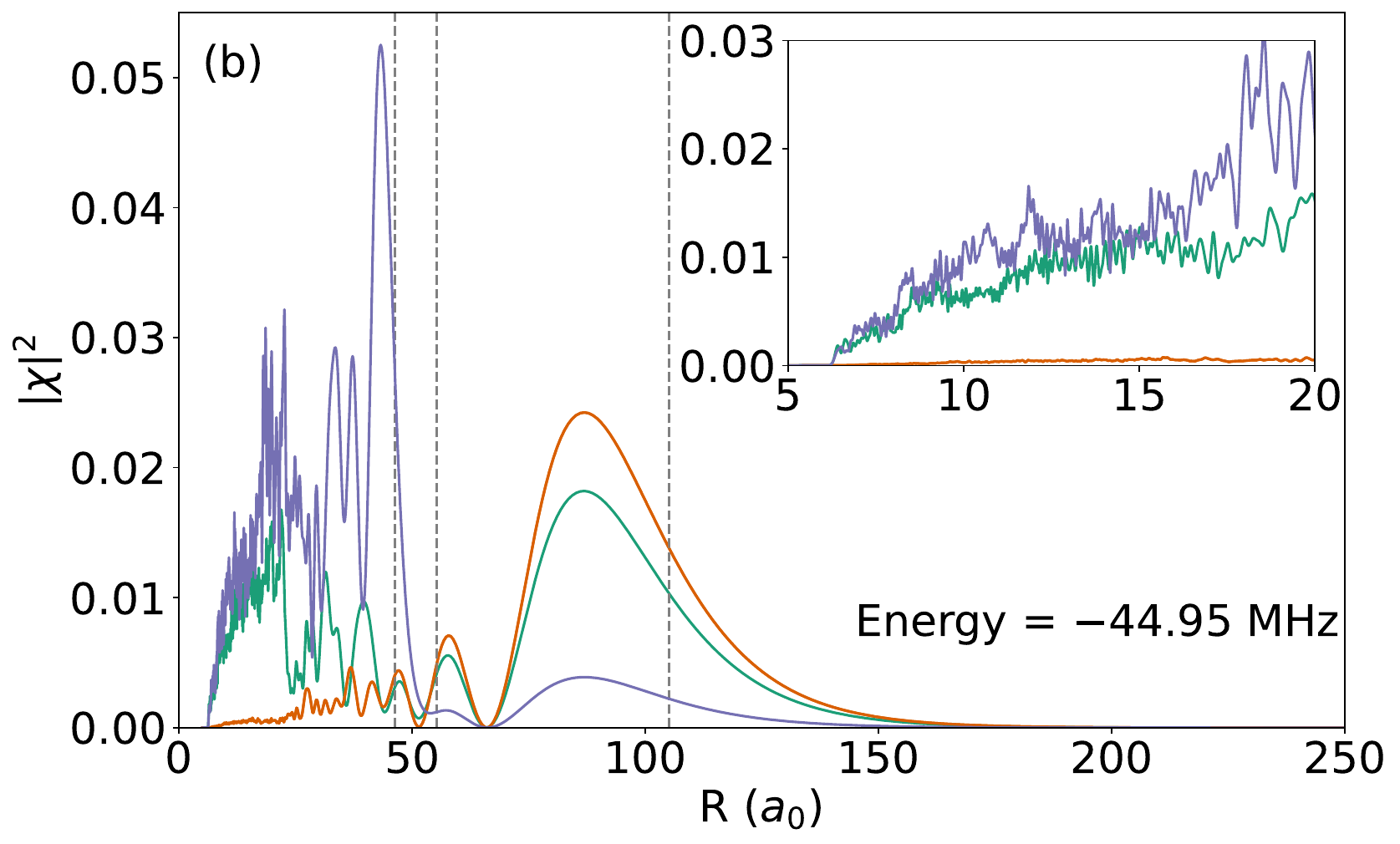}
\includegraphics[width=0.9\columnwidth]{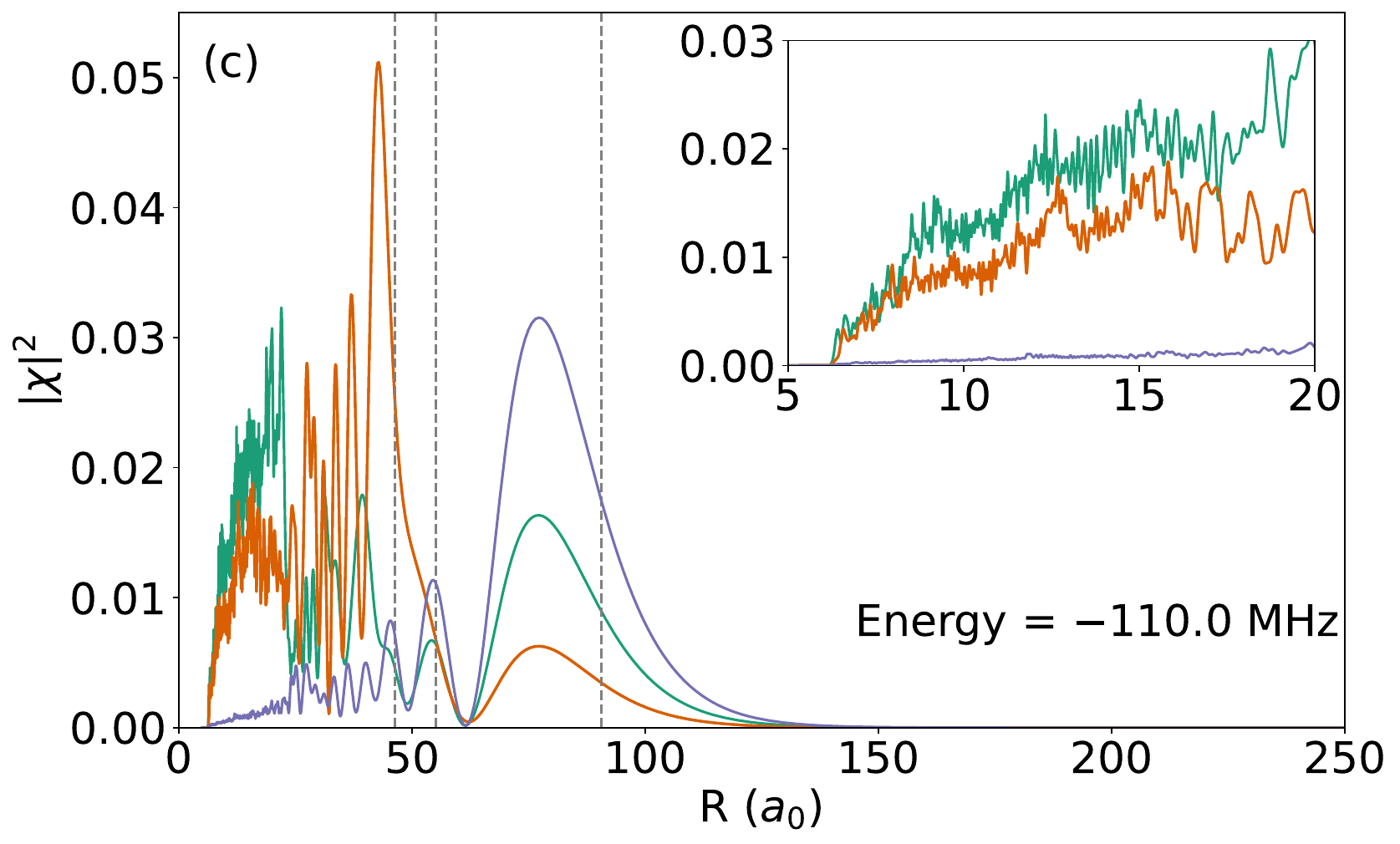}
\caption{Radial wavefunction densities for different near-threshold states at selected energies in the top bin: (a) $-10$~Mz; (b) $-44.95$~MHz; (c) $-110$~MHz. The states shown are those indicated with arrows in Fig.\ \ref{fig:scaling}. Scaling factors $\lambda$ are adjusted to place each state at the energy required for the panel concerned. The wavefunction densities are color-coded to match the states in Fig.\ \ref{fig:scaling}. The vertical dashed lines indicate the outer turning points for the lowest three adiabats at the corresponding energy. The insets show expanded views of the wavefunction densities at short range.}
\label{fig:wavefunctions}
\end{figure}

We wish to understand whether the slowly-varying states in Fig.\ \ref{fig:scaling} have true long-range character. To explore this, we calculate wavefunctions for the states indicated with arrows in Fig.\ \ref{fig:scaling} at three different energies within the top bin, shown by the horizontal lines at energies of $-10$, $-44.95$ and $-110$~MHz. The three states have node counts of 7255 (green), 7288 (orange) and 7320 (purple). At each energy, one of the states is slowly-varying and the other two are fast-varying. Each panel of Fig.\ \ref{fig:wavefunctions} shows radial wavefunction densities for the three states at the same energy. The vertical dashed lines on each panel indicate the outer turning points for the lowest three adiabats at the corresponding energy. The insets show expanded views of the wavefunction densities at short range.

\begin{table}[tb]
\caption{Short-range fractions $P_\textrm{short}$, defined as the integrated radial density for $0 < R <20\,a_0$, for each of the 3 states with wavefunctions shown in Fig.\ \ref{fig:wavefunctions}.}
\label{tab:frac-short}
\begin{ruledtabular}
\begin{tabular}{llll}
Energy / MHz & green & orange & purple
 \\ \hline
$-10$    & 0.0028 & 0.164  & 0.183  \\
$-44.95$ & 0.119  & 0.0053 & 0.175  \\
$-110$   & 0.229  & 0.152  & 0.0109 \\
\end{tabular}
\end{ruledtabular}
\end{table}

At each energy in Fig.\ \ref{fig:wavefunctions}, all three states have significant density around the outer turning point, with a point of inflection near the turning point. In each case it is the slowly-varying state in Fig.\ \ref{fig:scaling} that has the largest density in the long-range region. However, the biggest difference between the states is seen in their densities in the short-range region at $0 < R < 20\ a_0$, shown in the inset of each panel. This range is chosen because the levels of both K$_2$Rb and Li$_2$Yb have been shown to exhibit chaotic character for $\rho < 20\, a_0$ \cite{Croft:K+KRb:2017}. In each case the slowly-varying state has dramatically lower density in this region. The integrated densities $P_\textrm{short}$ for $0 < R < 20\,a_0$ are given in Table \ref{tab:frac-short}: the slowly-varying states have $P_\textrm{short} \lesssim 0.01$, while the fast-varying states all have $P_\textrm{short} > 0.1$. This demonstrates that the slowly-varying states have genuine long-range character, and are only weakly coupled to the chaotic bath of short-range states.

The avoided crossings visible in Fig.\ \ref{fig:scaling} show closest approaches between the slowly-varying state and the steep ones crossing it that are typically about 20 MHz, corresponding to coupling matrix elements around 10 MHz. This is significant on the scale of the bin depth, but is very small compared to the nearest-neighbor spacing for the chaotic short-range levels. This confirms that the slowly-varying (long-range) levels are not coupled strongly enough to the chaotic bath to form part of it.

\subsubsection{States in deeper bins}

\begin{figure}[tb]
\centering
\includegraphics[width=1.0\columnwidth]{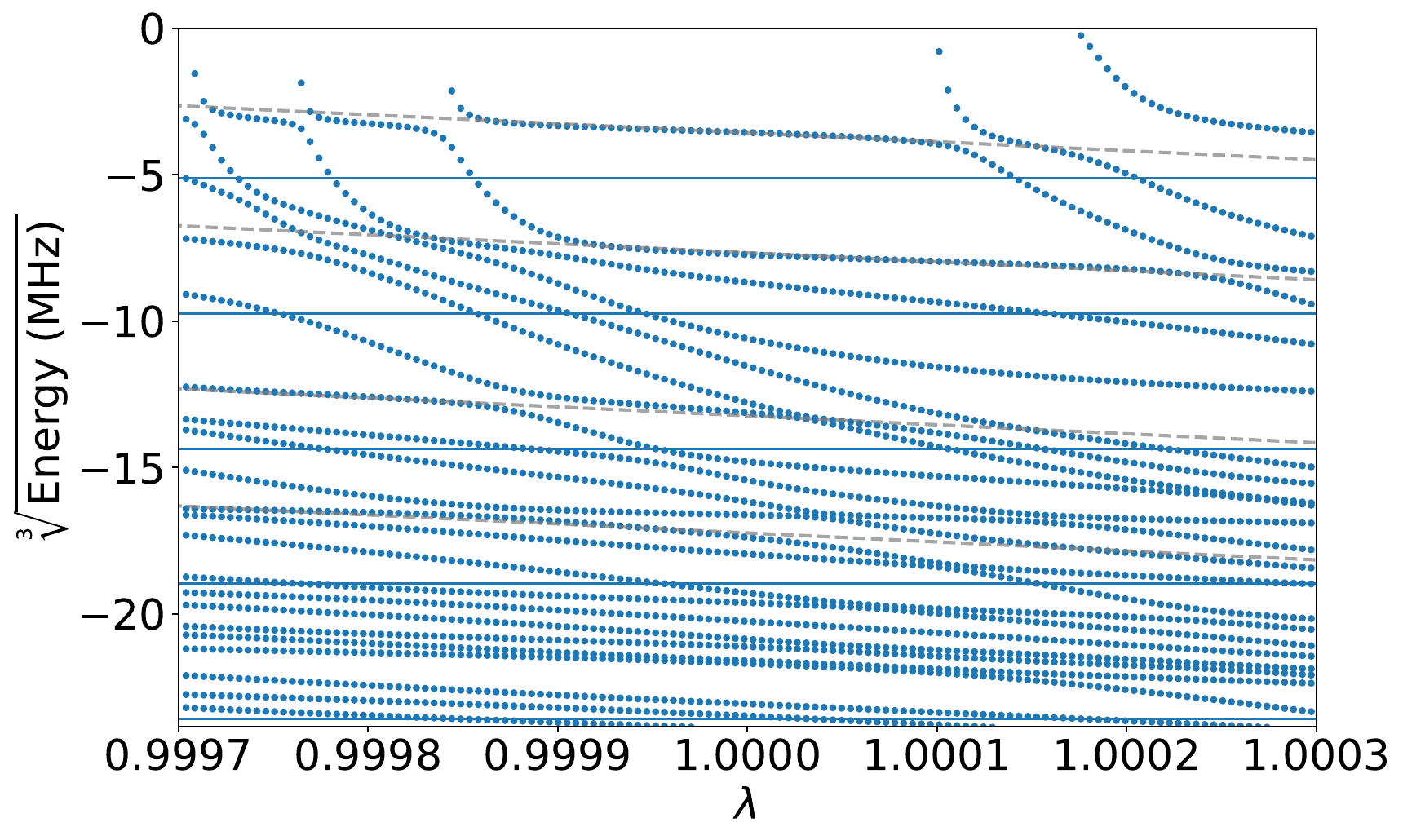}
\caption{Bound-state energies for states in the top 5 bins, as a function of the scaling parameter $\lambda$ for values close to $\lambda=1$. The grey lines show the general trend of the energies of states with long-range character. States close to the lines exist for many (but not all) values of $\lambda$.}
\label{fig:5bins-l1}
\end{figure}

An important question is whether the states with long-range character appear at similar positions in successive bins. This is the behavior expected for the long-range states in a single-channel problem \cite{LeRoy:1970, Gao:2000}, where the short-range part of the wavefunction varies very little with energy across the width of the top few bins. However, it might not be true in a system with chaotic structure at long range, as there the short-range part of the wavefunction might vary on the scale of the mean spacing $d$. To test this, Fig.\ \ref{fig:5bins-l1} shows the cube roots of binding energies of all states in the top 5 bins, as a function of $\lambda$ in the vicinity of $\lambda=1$. The near-threshold states are the ones whose energies vary most slowly with $\lambda$, with energies indicated with grey lines. Such states do not exist in every bin for every value of $\lambda$, because they undergo localized avoided crossings with other states. Nevertheless, the near-threshold states may be identified in most bins for most values of $\lambda$ in Fig.\ \ref{fig:5bins-l1}. Moreover, they are at approximately the same place within each bin, at least for the first 4 bins below threshold. This indicates that, in this system, the near-threshold states in different bins \emph{do} all share approximately the same ``quantum defect" defining their place within the bin.

It may be noted that the bin boundaries shown in Fig.\ \ref{fig:5bins-l1} are calculated for a pure $1/R^6$ potential. The real potential includes terms proportional to $1/R^8$ and higher. The effect of these higher-order terms is to increase the effective coefficient $C_6$ and to decrease the widths of successively deeper bins when plotted as the cube root of energy. This largely explains the apparent slow ``drift" of the near-threshold state to higher positions in deeper bins.

The near-threshold states are expected to become less distinct in deeper bins. As their binding energy increases, the inner turning point moves to lower $R$. The fraction of time spent at large $R$ decreases, so the coupling to short-range states increases. This effect may be seen in Fig.\ \ref{fig:5bins-l1}: the long-range states are clearly identifiable at most values of $\lambda$ in the top 3 bins, but are strongly perturbed in the 4th bin and are hard to identify with confidence in the 5th bin.

\begin{figure}[tb]
\centering
\includegraphics[width=1.0\columnwidth]{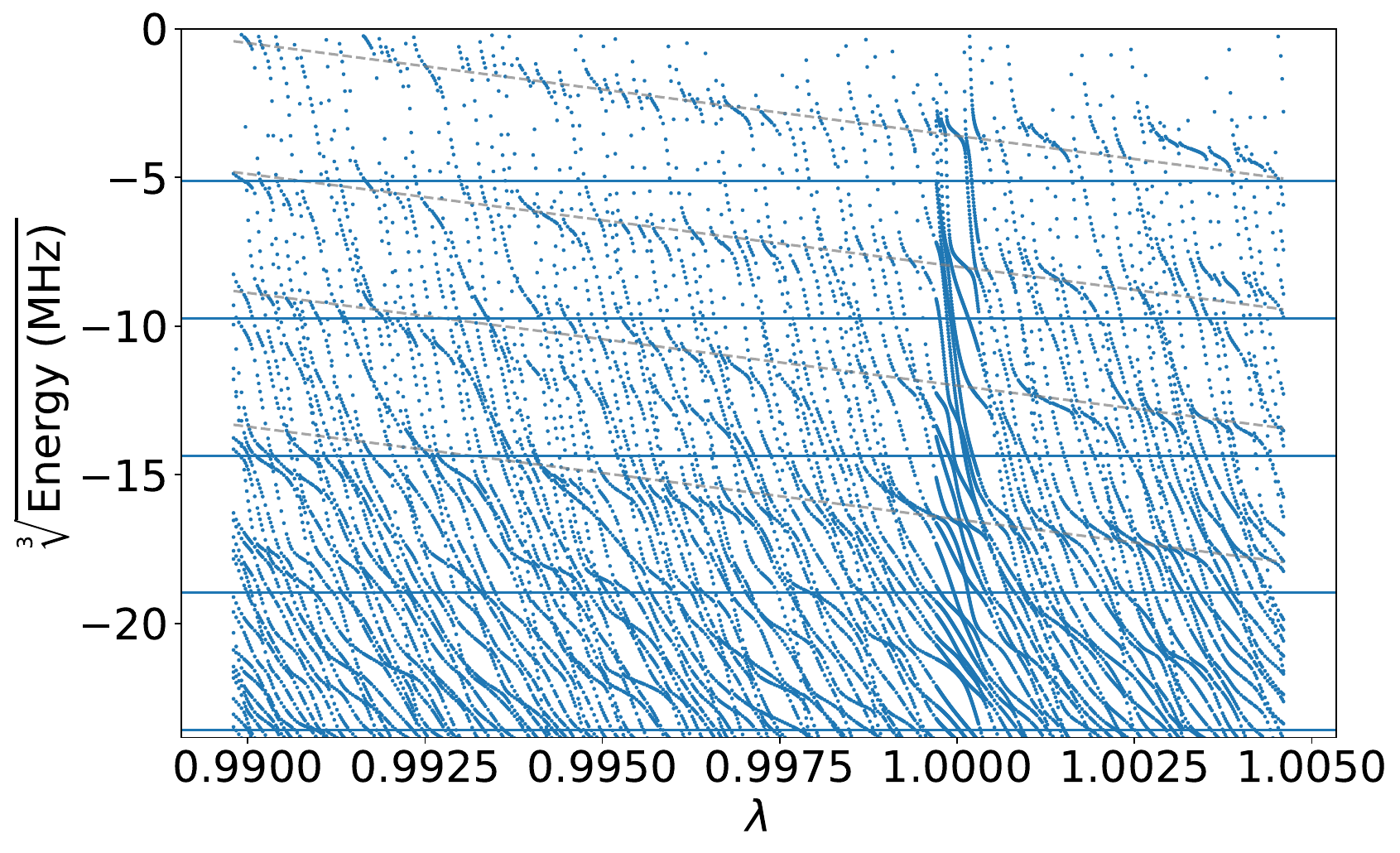}
\caption{Bound-state energies for states in the top 5 bins as a function of the scaling parameter $\lambda$, over a range of $\lambda$ sufficient for the long-range states to cross the full depth of each bin. The grey lines show the general trend of the energies of states with long-range character. These states undergo avoided crossings with shorter-range states, producing regions of low gradient of energy with respect to $\lambda$.}
\label{fig:5bins-wide}
\end{figure}

A wider scan of the near-threshold states is shown in Fig.\ \ref{fig:5bins-wide}. The trend lines of the long-range states are again indicated with grey lines. They are clear in bins 1 and 2 and visible in bins 3 and 4, but are hard to trace in bin 5. Since each real atom-diatom system has only one actual interaction potential, variations in behavior should be expected between systems. Some will have well-defined states with long-range character, particularly in the top bin, while in others the long-range states may be contaminated by accidental degeneracies with shorter-range states.

One particular limitation of the present calculations deserves mention. As a function of $\lambda$, states from many higher thresholds (with asymptotic $n>0$) cross threshold when $\lambda$ is varied enough to move a state with $n=0$ across a whole bin. However, magnetic Feshbach resonances are due to Zeeman perturbations, which are quite different from variations in $\lambda$. A magnetic field that is scanned over (say) 500 G will move thresholds for different spin states by about 1000 MHz relative to one another. Thus only states that lie within 1000 MHz of threshold can cause magnetic Feshbach resonances at such fields. Although \emph{many} higher thresholds produce steeply varying states that cross threshold as a function of $\lambda$ as in Figs.\ \ref{fig:5bins-l1} and \ref{fig:5bins-wide}, most of them lie too far below threshold (for any individual potential) to cause observable Feshbach resonances. Thus only a very few of the ``steeply varying" states in Figs.\ \ref{fig:5bins-l1} and \ref{fig:5bins-wide} will be important in ultracold collisions.

\begin{figure*}[tb]
\centering
	\subfloat[]{
		\includegraphics[width=0.45\textwidth]{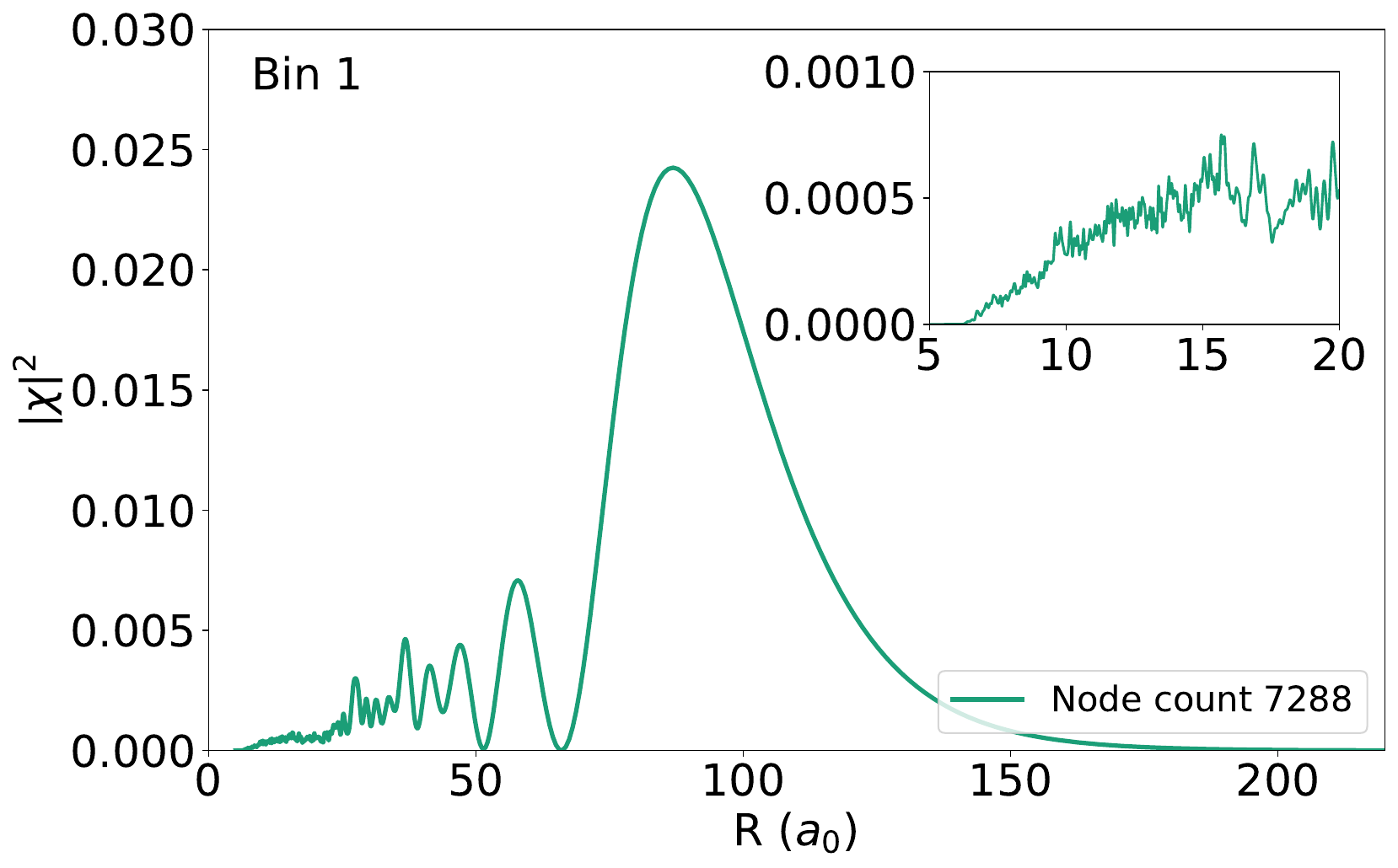}
	}
	\subfloat[]{
		\includegraphics[width=0.45\textwidth]{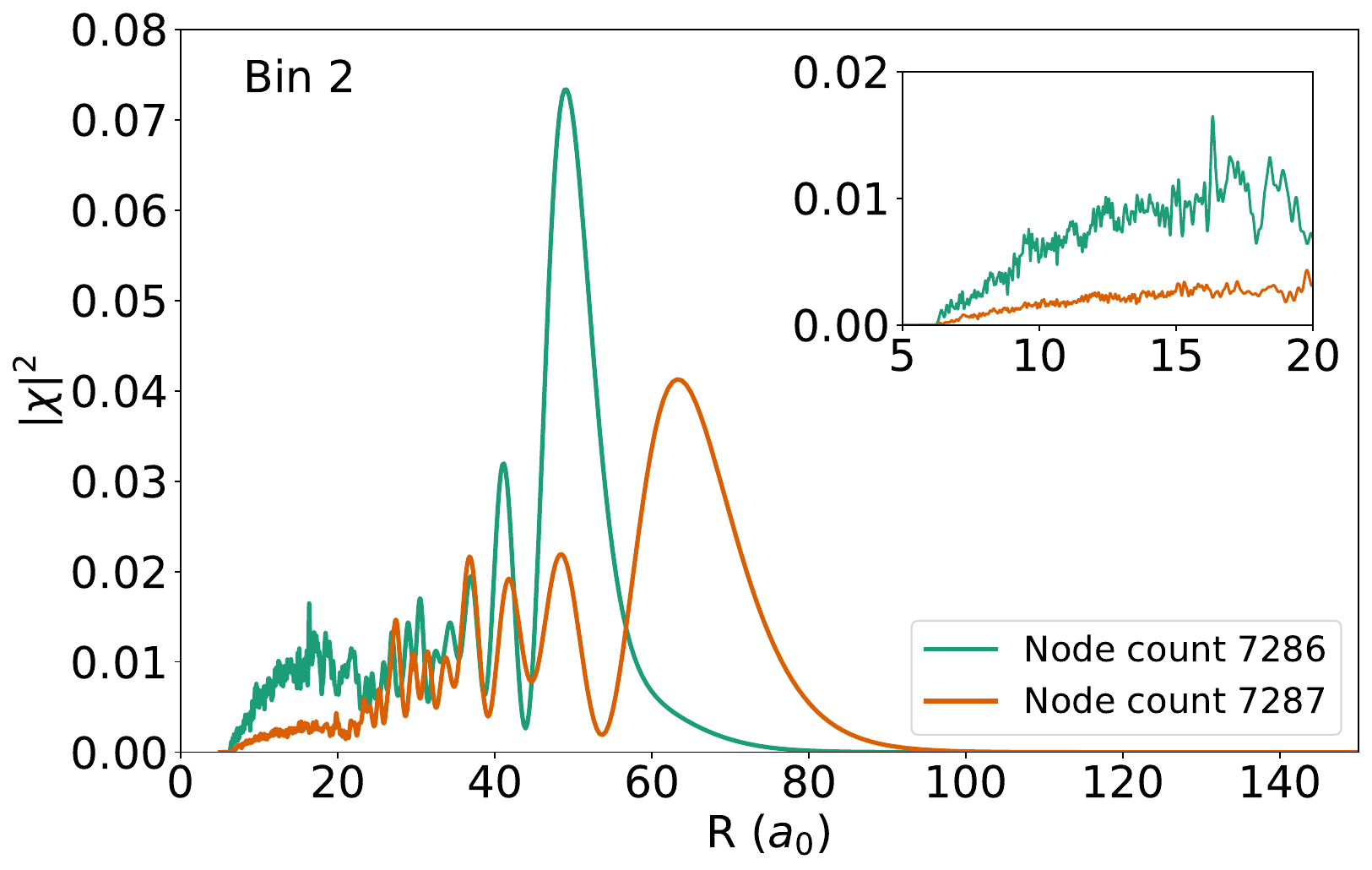}
	}
    \vspace{-0.8 cm}
	\subfloat[]{
		\includegraphics[width=0.45\textwidth]{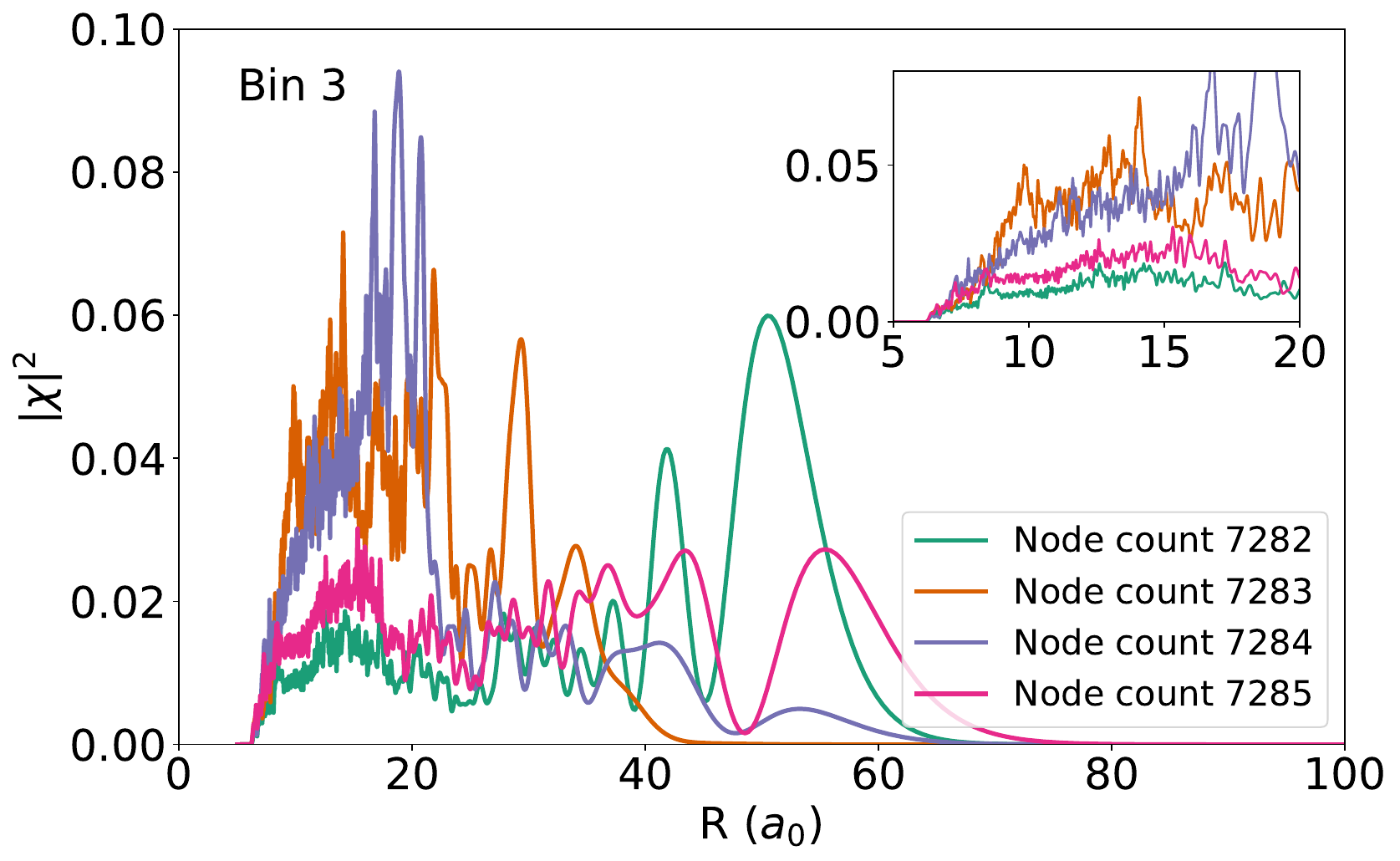}
	}
	\subfloat[]{
		\includegraphics[width=0.45\textwidth]{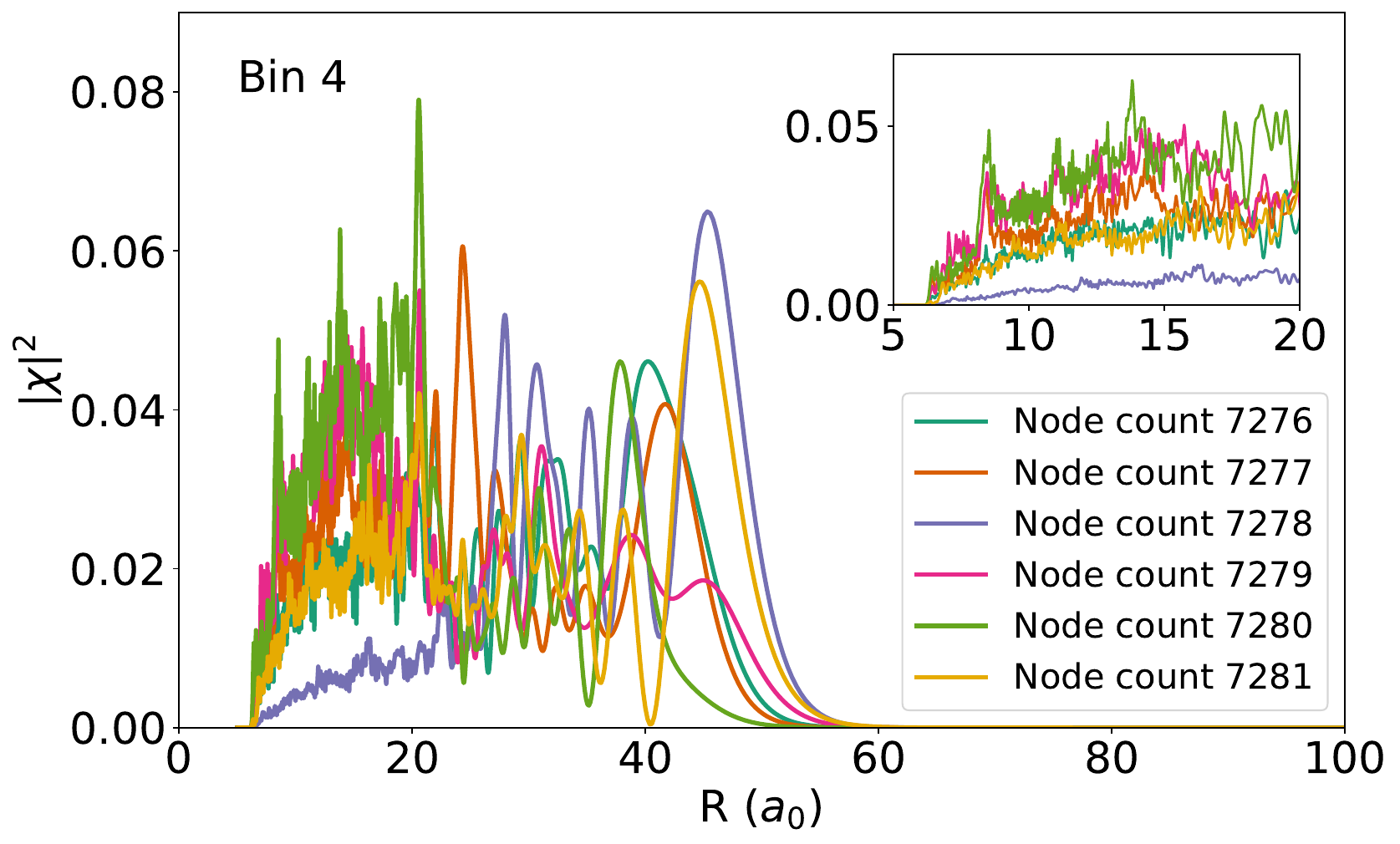}
	}
\caption{Radial wavefunction densities for all states in the top 4 bins for scaling factor $\lambda=1$. The insets show expanded views of the wavefunction densities at short range.}
\label{wavefunctions-top-4-bins}
\end{figure*}

To trace the progressive loss of long-range wavefunction character with binding energy, we have calculated wavefunctions for all states in the top 4 bins for the unscaled potential ($\lambda=1$). The resulting radial wavefunction densities are shown in Fig.\ \ref{wavefunctions-top-4-bins}. The one state in the top bin has clear long-range character, with density at $R<20\ a_0$ that is only about 1\% of the maximum. There are two states in bin 2; the extra one must be due to a higher threshold. One of them (orange in Fig.\ \ref{wavefunctions-top-4-bins}) clearly has much more long-range character than the other, but the difference in short-range density between them is only about a factor of 4. This is probably because the second state (blue) arises mostly from $n=1$, whose outer turning point at this energy is near 53 $a_0$, so that there is significant mixing between the two states at fairly long range. There are 4 states in bin 3; one of them (red in Fig.\ \ref{wavefunctions-top-4-bins}) clearly has more density at long range, but now there is only a factor of 2 to 3 difference in the densities at short range. The situation is similar in bin 4, where one state (green) still has much lower short-range density, by a factor of around 3.

The short-range densities seen in Fig.\ \ref{wavefunctions-top-4-bins} accord with the expected behavior of long-range states \cite{LeRoy:1970}. For an interaction potential of the form $-C_6/R^6$, the short-range wavefunction amplitude of a long-range state with binding energy $E$ (relative to the threshold that supports it) is proportional to $|E|^{1/3}$ \cite{Brue:AlkYb:2013}. The amplitude thus increases approximately linearly with bin number, and the short-range radial density increases quadratically. The couplings between long-range and (chaotic) short-range states will increase approximately linearly with bin number, and will not approach the density of short-range states until at least bin 10, which lies more than 100 GHz below threshold. The long-range states are thus unlikely to be coupled strongly enough to the chaotic bath of short-range states to become part of it for at least this depth below threshold.

The overall picture that arises from these results is that states of strongly long-range character can exist, even in systems that exhibit chaotic behavior at short range. Long-range states that are near the top of the well are unlikely to be perturbed much by the short-range states. For Rb+KRb, this applies strongly to states in the top few bins (within a few GHz of threshold), but is likely to persist much deeper, until eventually the couplings to short-range states are sufficient that the long-range states lose their identity. Despite this, there is always a possibility that (for the real, unknown interaction potential) a long-range state will experience an accidental near-degeneracy with a shorter-range one that results in significant mixing between them; when this happens, there may be no \emph{individual} state that has strong long-range character. This might explain the wide variation in collisional loss rates observed for different atomic hyperfine states in $^{39}$K+Na$^{39}$K~\cite{Voges:2022}.

\section{Conclusions}

We have investigated the near-threshold states that may exist in systems with deep potential wells and strong coupling at short range, with chaotic dynamics for short-range states. We considered a simplified model of Rb+KRb that neglects electron and nuclear spins. We carried out bound-state coupled-channel calculations in full dimensionality using hyperspherical methods, and in reduced dimensionality using atom-diatom methods. We encountered convergence problems in the hyperspherical calculations, but nevertheless used them to provide accurate densities of states as a function of energy. These densities of states demonstrated that there are large numbers of additional states that exist close to threshold when the long-range tail of the potential is fully taken into account.

The hyperspherical and atom-diatom approaches produce sets of adiabats with quite similar features. Furthermore, energy levels from atom-diatom calculations show clear signatures of quantum chaos at energies about two-thirds of the way up the potential well. Specifically, the levels show strong level repulsion and their nearest-neighbor spacings accurately follow Wigner-Dyson statistics in this region. However, the extent of chaos decreases substantially higher up the well, closer to dissociation (i.e.\ threshold). This is as expected if there are long-range states with regular behavior that are only weakly coupled to the chaotic states that exist at short range.

We next used coupled-channel calculations in the atom-diatom representation to find all bound states that exist within about 14 GHz of threshold. We repeated this calculation for 696 different potentials, obtained from the original by multiplying by a scaling factor $\lambda$ in the range $0.99 < \lambda < 1.005$. These scans showed persistent states with long-range character, crossing and avoided-crossing with many short-range states as a function of $\lambda$. For a particular interaction potential, the long-range states might be significantly perturbed by interaction with a single shorter-range state. However, the strengths of the avoided crossings were far too low (and thus the couplings between long-range and short-range states were far to weak) for the long-range states to become part of the chaotic bath.

Extrapolating the couplings using long-range theory suggests that states with long-range character will persist until at least 100 GHz below threshold. This is sufficient that, at the energy of the lowest threshold, there will be states with long-range character supported by multiple rotationally excited (and hyperfine-excited) thresholds. This is the requirement for the appearance of magnetically tunable Feshbach resonances due to long-range states.

The long-range states are not chaotic in character, and there is no reason to expect their widths and lifetimes to follow RRKM theory and be determined by the density of states. Instead, their widths will be determined by the strengths of interchannel couplings; these are weak at long range, so that the states may have very long lifetimes even when inelastic loss channels are available.

This work provides an overall picture of the near-threshold states that exist for systems with strong short-range coupling, such as the alkali atom-molecule and molecule-molecule systems. There are sets of regular long-range states just below each scattering threshold, superimposed on chaotic manifolds of short-range states. The long-range states have low probability densities at short range, so they are only weakly coupled to the short-range states. The interaction potential is weakly anisotropic at long range, so the long-range states will show slow inelastic decay even when there are open channels. They will also have low rates of laser-induced loss, which can occur only at short range. When the thresholds can be tuned with respect to one another, for example with magnetic fields for atom-diatom complexes, the long-range states can produce narrow magnetically tunable Feshbach resonances.

The short-range states, by contrast, are strongly coupled to one another. They are chaotic in nature; when they are above threshold, they probably have lifetimes determined by RRKM theory and proportional to the density of states. They are also subject to fast laser-induced loss, which may increase the widths further. They may cause fast collisional loss when one of them is close to the energy of the incoming channel. However, the resonances due to short-range states do not necessarily overlap, so that different systems (and different hyperfine states of the same system) may show differing behaviour.

There is no sharp distinction between long-range and short-range states, because the mixing between them increases with the binding energy of the long-range state below the threshold that supports it. Far enough below threshold, the long-range states merge into the chaotic bath. Nevertheless, there is a substantial range of energies below threshold where distinct long-range states are likely to exist.

The existence of long-range states may explain the long lifetimes observed for collision complexes in ultracold systems \cite{Gersema:2021, Nichols:long-lived:2021} and the existence of narrow Feshbach resonances in systems such as $^{40}$K+Na$^{40}$K \cite{Yang:K_NaK:2019, Frye:long-range:2023} and $^{39}$K+Na$^{39}$K \cite{Meyer:2026}. The weak coupling between the short-range and long-range states may explain how such resonances can exist in the presence of fast background loss.

\section*{Rights retention statement}

For the purpose of open access, the authors have applied a Creative Commons Attribution (CC BY) licence to any Author Accepted Manuscript version arising from this submission.

\section*{Data availability statement}

The data that support the findings of this article are openly available at doi:10.15128/r1m326m1814.

\section*{Acknowledgement}
We are grateful to Matthew Frye and Ruth Le Sueur for valuable discussions.
This work was supported by the U.K. Engineering and Physical Sciences Research Council (EPSRC) Grant Nos.\
EP/P01058X/1, 
EP/W00299X/1, 
EP/Z535898/1  
and UKRI2226. 
B.K.K. acknowledges that part of this work was done under the auspices of the US Department of Energy under Project No.\ 20240256ER of the Laboratory Directed Research and Development Program at Los Alamos National Laboratory. This work used resources provided by the Los Alamos National Laboratory Institutional Computing Program. Los Alamos National Laboratory is operated by Triad National Security, LLC, for the National Nuclear Security Administration of the U.S. Department of Energy (contract No.\ 89233218CNA000001).

\bibliographystyle{long_bib}
\bibliography{../all.bib}

\end{document}